\documentclass[sigconf]{acmart}
\AtBeginDocument{%
  \providecommand\BibTeX{{%
    \normalfont B\kern-0.5em{\scshape i\kern-0.25em b}\kern-0.8em\TeX}}}

\acmConference[SIGMOD '22]{Proceedings of the 2022 International Conference on Management of Data}{June 12--17, 2022}{Philadelphia, PA, USA.}
\acmBooktitle{Proceedings of the 2022 International Conference on Management of Data (SIGMOD '22), June 12--17, 2022, Philadelphia, PA, USA}
\acmPrice{15.00}
\acmISBN{978-1-4503-XXXX-X/18/06}

\usepackage{multirow}
\begin{document}
\title{LOCAT: Low-Overhead Online Configuration Auto-Tuning of Spark SQL Applications}

\author{Jinhan Xin}
\email{jh.xin@siat.ac.cn}
\orcid{0000-0003-1900-7774}
\affiliation{%
  \institution{Shenzhen Institute of Advanced Technology (SIAT), Chinese Academy of Sciences (CAS)}
  \streetaddress{1068 Xueyuan Avenue, Shenzhen University Town}
  \city{Shenzhen}
  \state{Guangdong}
  \country{China}
  \postcode{518055}
}

\affiliation{%
  \institution{University of Chinese Academy of Sciences (UCAS)}
  \streetaddress{19 Yuquan Road, Shijinshan District}
  \city{Beijing}
  \country{China}
  \postcode{100049}
}

\author{Kai Hwang}
\email{hwangkai@cuhk.edu.cn}
\orcid{0000-0001-5503-3932}
\affiliation{%
  \institution{The Chinese University of Hong Kong, Shenzhen}
  \streetaddress{2001 Longxiang Boulevard, Longgang District}
  \city{Shenzhen}
  \state{Guangdong}
  \country{China}
  \postcode{518172}
 }
 
 \author{Zhibin Yu}
\authornote{Zhibin Yu is the corresponding author.}
\email{zb.yu@siat.ac.cn}
\orcid{0000-0001-8067-9612}
\affiliation{%
  \institution{Shenzhen Institute of Advanced Technology (SIAT), Chinese Academy of Sciences (CAS)}
  \streetaddress{1068 Xueyuan Avenue, Shenzhen University Town}
  \city{Shenzhen}
  \state{Guangdong}
  \country{China}
  \postcode{518055}
}

\affiliation{%
  \institution{Shuhai Lab, Shenzhen Huawei Cloud Co.,Ltd.}
  \streetaddress{Bantian, Longgang District}
  \city{Shenzhen}
  \state{Guangdong}
  \country{China}
  \postcode{518129}
}

\renewcommand{\shortauthors}{Jinhan Xin, Kai Hwang, and Zhibin Yu.}

\begin{abstract}
  Spark SQL has been widely deployed in industry but it is challenging to tune its performance. Recent studies try to employ machine learning (ML) to solve this problem. They however suffer from two drawbacks. First, it takes a long time (high overhead) to collect training samples. Second, the optimal configuration for one input data size of the same application might not be optimal for others. 

To address these issues, we propose a novel Bayesian Optimization (BO) based approach named LOCAT to automatically tune the configurations of Spark SQL applications online. LOCAT innovates three techniques. The first technique, named QCSA, eliminates the configuration-insensitive queries by Query Configuration Sensitivity Analysis (QCSA) when collecting training samples. The second technique, dubbed DAGP, is a Datasize-Aware Gaussian Process (DAGP) which models the performance of an application as a distribution of functions of configuration parameters as well as input data size. The third technique, called IICP, Identifies Important Configuration Parameters (IICP) with respect to performance and only tunes the important parameters. As such, LOCAT can tune the configurations of a Spark SQL application with low overhead and adapt to different input data sizes. 

We employ Spark SQL applications from benchmark suites $TPC-DS$, $TPC-H$, and $HiBench$ running on two significantly different clusters, a four-node ARM cluster and an eight-node x86 cluster, to evaluate LOCAT. The experimental results on the ARM cluster show that LOCAT accelerates the optimization procedures of Tuneful~\cite{fekry2020tuneful}, DAC~\cite{yu2018datasize}, GBO-RL~\cite{kunjir2020black}, and QTune~\cite{li2019qtune} by factors of $6.4\times$, $7.0\times$, $4.1\times$, and $9.7\times$ on average, respectively. On the x86 cluster, LOCAT reduces the optimization time of Tuneful, DAC, GBO-RL, and QTune by factors of $6.4\times$, $6.3\times$, $4.0\times$, and $9.2\times$ on average, respectively. Moreover, LOCAT improves the performance of the applications on the ARM cluster tuned by Tuneful, DAC, GBO-RL, and QTune by factors of $2.4\times$, $2.2\times$, $2.0\times$, and $1.9\times$ on average, respectively. On the x86 cluster, LOCAT improves the performance of these applications tuned by Tuneful, DAC, GBO-RL, and QTune by factors of $2.8\times$, $2.6\times$, $2.3\times$, and $2.1\times$ on average, respectively.

\end{abstract}




\maketitle

\section{Introduction}\label{sec:intro}

Big data query systems such as Hive~\cite{hive}, Presto~\cite{sethi2019presto}, and Spark SQL~\cite{armbrust2015spark} have been widely deployed in industry to mine valued information from massive data efficiently~\cite{qayyum2020roadmap}. As a higher level library on top of Apache Spark~\cite{spark-original}, Spark SQL not only inherits Spark's excellent big data processing capabilities, but also provides support for query-like large-scale data analysis, such as OnLine Analytical Processing (OLAP)~\cite{lv2015olap, cuzzocrea2015data}.

However, it is challenging to tune the configuration parameters of a Spark SQL application for optimal performance because of two reasons. First, the lower layer Spark and the upper layer Spark SQL both have a number (e.g.,$>20$) of configuration parameters. Not only the ones for Spark SQL (e.g., {\it spark.sql.shuffle.partitions}) itself, but also those for Spark (e.g., {\it spark.executor.memory}) significantly affect the performance of a Spark SQL application. For example, the parameter {\it spark.executor.memory} specifies the amount of memory used by an executor process~\cite{SparkConfWebsite}. Too large value of it may cause a long garbage collection time~\cite{chiba2016workload} pausing the application whereas too small value may even lead to out of memory (OOM) errors~\cite{kunjir2020black}. Therefore, tuning the larger number of parameters for optimal performance of a Spark SQL application is difficult. Second, the configuration parameters within the same layer as well as from different layers may intertwine with each other in a complex way with respect to performance, further troubling the performance tuning of a Spark SQL application.

Recent studies propose to leverage machine learning (ML) to tune the configurations for Spark programs~\cite{yu2018datasize, kunjir2020black} and database systems~\cite{li2019qtune, zhang2019end}. However, current ML-based approaches have two drawbacks. \textbf{First}, these approaches take a long time to collect training samples, which is inconvenient in practice. The long time stems from four factors. (1) The number of training samples is large (e.g., 1000 --- 10000), which is the nature of ML-based approaches. (2) The time used to collect each training sample of an application is typically long (e.g., several minutes) because it is collected by running the application on a real cluster with a random configuration. (3) A Spark SQL application typically consists of a number (e.g., 20) of queries. The more queries in a Spark SQL application generally make it take longer time to execute and in turn longer time to collect one training sample. (4) ML-based approaches generally need more training samples when tuning more parameters. 

\textbf{Second}, most ML-based approaches can not adapt to the changes of input data sizes of a Spark SQL application. That is, a configuration making a Spark SQL application achieve the optimal performance for one input data size might not produce optimal performance for another input data size. This makes the same application need to be re-tuned when its input data size is changed, which is time-consuming. However, customers typically do not change their Spark SQL applications frequently while definitely change  the input data size of the same application.

To address these issues, we propose a novel approach dubbed LOCAT to automatically tune the configurations of a Spark SQL application online. LOCAT's first key innovation is that we observe an important as well as interesting finding: {\it different queries in a Spark SQL application respond to configuration parameter tuning with significantly different sensitivity}. Some queries of an application are insensitive to the parameter tuning at all, and we therefore call them {\it configuration-insensitive} queries. Based on this finding, we remove the configuration-insensitive queries from a Spark SQL application when we run the application with random configurations to collect training samples. As such, the sample collection time can be dramatically reduced. 

\begin{figure}
  \centering
  \includegraphics[width=\linewidth]{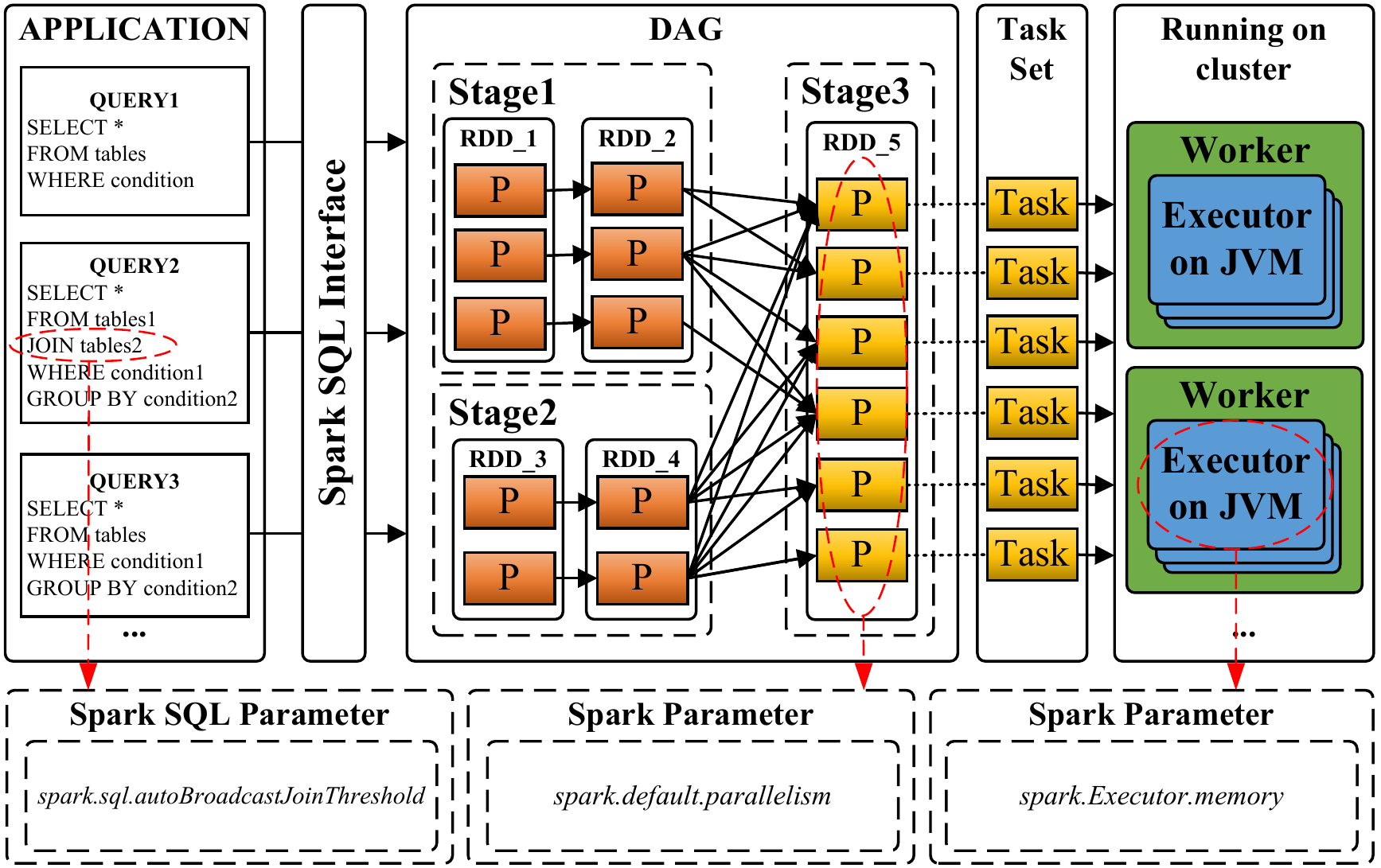}
  \vspace{-4mm}
  \caption{An Overview of the Spark SQL framework.}
  \label{fig:duck1}
  \vspace{-6mm}
\end{figure}

The second key innovation is that we propose a Datasize-Aware Gaussian Process (DAGP) to take the input data size in addition to the configuration parameters of a Spark SQL application into consideration as tuning the configuration parameters. In contrast, other Gaussian Process (GP) based approaches such as CherryPick~\cite{alipourfard2017cherrypick} only consider the configuration parameters, which needs to perform the time-consuming parameter re-tuning when an application's input data size is changed. The third innovation is that we propose to identify the important configuration parameters of a Spark SQL application and in turn only tune them in BO (Bayesian Optimization) iterations. We name this technique IICP. Generally, tuning more parameters takes more iterations to find the optimal configuration for an application by BO. IICP therefore takes less iterations and in turn shorter time to find the optimal configuration.

In particular, this paper makes the following contributions.
\begin{itemize}
    \item We find that some queries of a Spark SQL application are insensitive to configuration parameter tuning with respect to performance by Query Configuration Sensitivity Analysis (QCSA).
    \item We propose to leverage Gaussian Process (GP) to model the relationship between performance and the input data size of a Spark SQL application in addition to the configuration parameters. As such, our approach can adapt to different input data sizes of the same application and we name this technique DAGP (Datasize-Aware Gaussian Process).
    \item We propose to identify the important configuration parameters (IICP) of a Spark SQL application and only tune these parameters in order to reduce the tuning time.
    \item By putting it all together, we develop an online configuration parameter tuning approach for Spark SQL applications with low overhead, named LOCAT.
   \item We employ Spark SQL applications from benchmark suites $TPC-DS$, $TPC-H$, and $HiBench$ running on two different clusters ---  a four-node ARM cluster and an eight-node x86 cluster --- to evaluate LOCAT. The results on the ARM cluster show that LOCAT accelerates the optimization procedures of Tuneful~\cite{fekry2020tuneful}, DAC~\cite{yu2018datasize}, GBO-RL~\cite{kunjir2020black}, and QTune~\cite{li2019qtune} by factors of $6.4\times$, $7.0\times$, $4.1\times$, and $9.7\times$ on average, respectively. On the x86 cluster, LOCAT reduces the optimization time used by Tuneful, DAC, GBO-RL, and QTune by factors of $6.4\times$, $6.3\times$, $4.0\times$, and $9.2\times$ on average, respectively. Moreover, LOCAT improves the applications' performance on the ARM cluster tuned by Tuneful, DAC, GBO-RL, and QTune by factors of $2.4\times$, $2.2\times$, $2.0\times$, and $1.9\times$ on average, respectively. On the x86 cluster, LOCAT improves the applications' performance tuned by Tuneful, DAC, GBO-RL, and QTune by factors of $2.8\times$, $2.6\times$, $2.3\times$, and $2.1\times$ on average, respectively.
\end{itemize}

The rest of this paper is organized as follows. Section~\ref{sec:bkg} introduces the background and motivation. Section~\ref{sec:locate} presents our LOCAT approach. Section~\ref{sec:exp} describes the experimental setup. Section~\ref{sec:rlt} provides and analyzes the experimental results. Section~\ref{sec:rlw} describes the related work and Section~\ref{sec:con} concludes the paper.

\section{Background and Motivation}\label{sec:bkg}

\subsection{Spark SQL Framework}\label{sparksqlback}
Spark SQL~\cite{armbrust2015spark} is built on top of Apache Spark~\cite{spark-original} to facilitate high-performance structured data processing. Unlike Spark RDD APIs, Spark SQL interfaces provide Spark with more information about the structure of both data and computation being performed~\cite{armbrust2015spark}. It is therefore widely used in industry~\cite{baldacci2018cost} such as OLAP~\cite{lv2015olap}. As shown in Figure~\ref{fig:duck1}, a Spark SQL application typically consists of a number of queries. The Spark SQL framework transforms each query into a DAG (directed acyclic graph) which is then split into a collection of stages consisting of a set of parallel tasks. Each task corresponds to a partition (P in Figure~\ref{fig:duck1}) computing partial results of an application. Each stage may depend on other stages, called lineage stored in a RDD. 

The DAG scheduler of Spark schedules the tasks on several executors to execute in parallel. This parallelism is controlled by several {\it configuration parameters}. For example, in Yarn~\cite{vavilapalli2013yarn} mode, the parameter {\it spark.executor.instances} specifies the number of executors, and {\it spark.executor.cores} specifies the number of cores used by each executor. The product of the number of executors and the number of cores per executor determines the maximum number of total tasks that can be executed by a Spark SQL cluster at a time.

In summary, the performance of a Spark SQL application is controlled by more than 200 configuration parameters, which can be generally classified into two levels: the Spark SQL configuration parameters (upper level) and the Spark core ones (lower level). The upper level parameters specify the properties of a Spark SQL application. For example, {\it spark.sql.autoBroadcastJoinThreshold} specifies the maximum size in bytes for a table that is broadcasted to all workers when performing a {\it join} operation, which significantly affects its performance. The lower level parameters specify fourteen aspects of the Spark core such as {\it execution parallelism} and {\it memory management}. For instance, the above mentioned {\it spark.executor.cores} and {\it spark.executor.instances} control the computing parallelism, and dramatically influence the performance of a Spark SQL application too. Moreover, the upper level configurations may interact with the lower level ones in a complex way, which makes tuning configurations for a Spark SQL application extremely difficult.

\subsection{Bayesian Optimization}

Bayesian Optimization (BO)~\cite{mockus2012bayesian} is a principled technique based on Bayes Theorem to direct an efficient and effective search of a global optimization problem. It minimizes or maximizes an objective function {\it f} iteratively through adaptive sampling of the search space with a manner that balances exploration and exploitation. BO has two key components: {\it surrogate model} and {\it acquisition function}. The surrogate model is used to model the objective function {\it f} and the acquisition function guides the selection of the next evaluation sample. BO iteratively fits the surrogate model by using the samples selected by the acquisition function and finally finds the minimal/maximal objective function {\it f}.

The surrogate models can be other machine learning models such as Random Forest (RF) and Boosted Regression Trees (BRT) that have a good ability to model the non-linear interactions~\cite{hsu2018arrow}. However, they are weak in theoretical guarantees on the confidence bounds while GP (Gaussian Process) isn't~\cite{kunjir2020black}. Moreover, GP has many outstanding features such as supporting for noisy observations and gradient-based methods~\cite{shahriari2015taking}. We therefore apply GP~\cite{rasmussen2003gaussian,williams1998bayesian} as the surrogate model of BO in this work. 

As for acquisition functions, the popular ones are expected improvement (EI)~\cite{jones1998efficient}, probability of improvement (PI)~\cite{hoffman2011portfolio}, and GP upper confidence bound (GP-UCB)~\cite{rasmussen2003gaussian}. Among which, EI is one of the most widely used acquisition functions.
However, we do not directly use EI. Instead, we leverage the EI with Markov Chain Monte Carlo (EI-MCMC)~\cite{snoek2012practical} for better overall performance, which has shown better performance compared to other acquisition functions across a wide range of test cases~\cite{snoek2012practical}.

\subsection{Motivation}
 
\begin{figure}[t!]
 \centering
 \includegraphics[width=\linewidth]{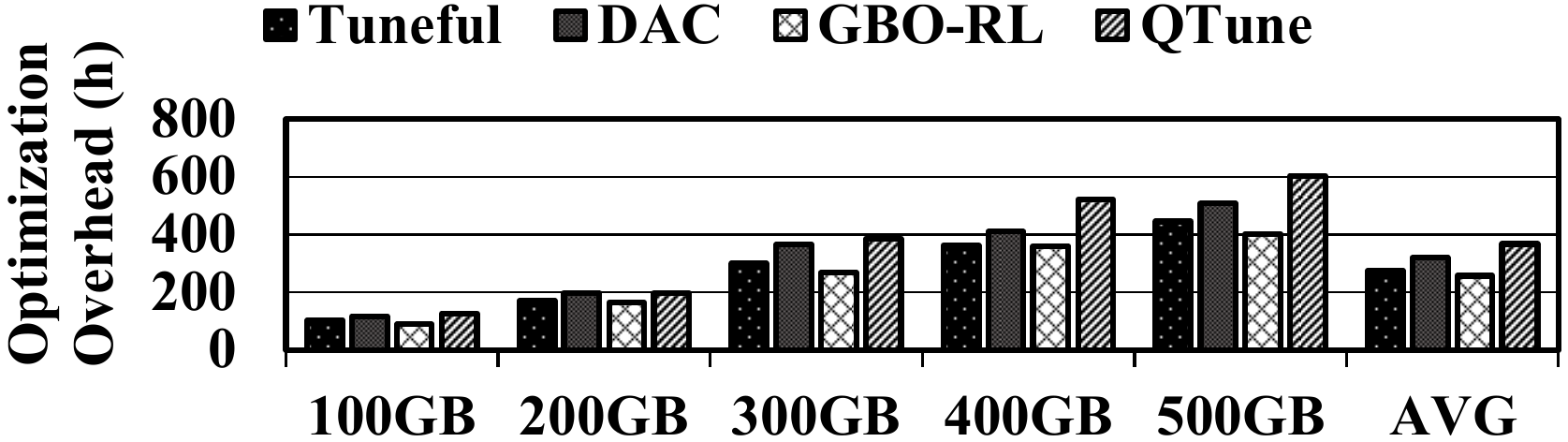}
 \vspace{-6mm}
 \caption{The time used to find the optimal configuration of $TPC-DS$ by Tuneful, DAC, GBO-RL, and QTune.}
 \label{fig:ovhd}
 \vspace{-4mm}
\end{figure}
Although the time (e.g., seconds or minutes) used to execute a Spark SQL application is shorter than that (e.g., hours or days) used to optimize its execution, it is still necessary to optimize it because the optimization is a one-shot task while application is repeatedly executed many times in a long time such as months or years. Saving a short time in each execution would accumulate a long time, which is a large benefit. To optimize Spark SQL applications, the easiest way is to employ the state-of-the-art (SOTA) approaches. However, we find that these approaches all take a long time (e.g., days or weeks) to find the optimal configuration. Figure~\ref{fig:ovhd} shows the time used by four SOTA approaches (Tuneful, GBO-RL, DAC for Spark applications, and QTune for database systems) to find the optimal configuration for $TPC-DS$. We made two observations. For one, the time used by these approaches is at least 
89 hours (GBO-RL) when the input data size of $TPC-DS$ is 100 GB. Second, the time used by all the approaches is getting significantly longer when the input data size becomes larger.  

In industry, the input data size of a Spark SQL application is typically from hundreds of Giga bytes to Tera bytes, even Peta bytes~\cite{qayyum2020roadmap}. In such a case, it is very inconvenient, if feasible, to apply the above approaches to find the optimal configurations for Spark SQL applications. For example, we optimize $TPC-DS$ with 500$\:$GB of data by using GBO-RL on our ARM cluster, it took 402 hours (16.75 days)! This motivates this work. 

\section{LOCAT Approach}\label{sec:locate}
\subsection{Overview}
LOCAT is a configuration auto-tuning approach that automatically finds the optimal values of configuration parameters for an application running on a given cluster in a short time. It is designed for a common industrial usage: a Spark SQL application repeatedly runs many times with the size of input data changing over time.

\begin{figure}[t!]
  \centering
  \includegraphics[width=\linewidth]{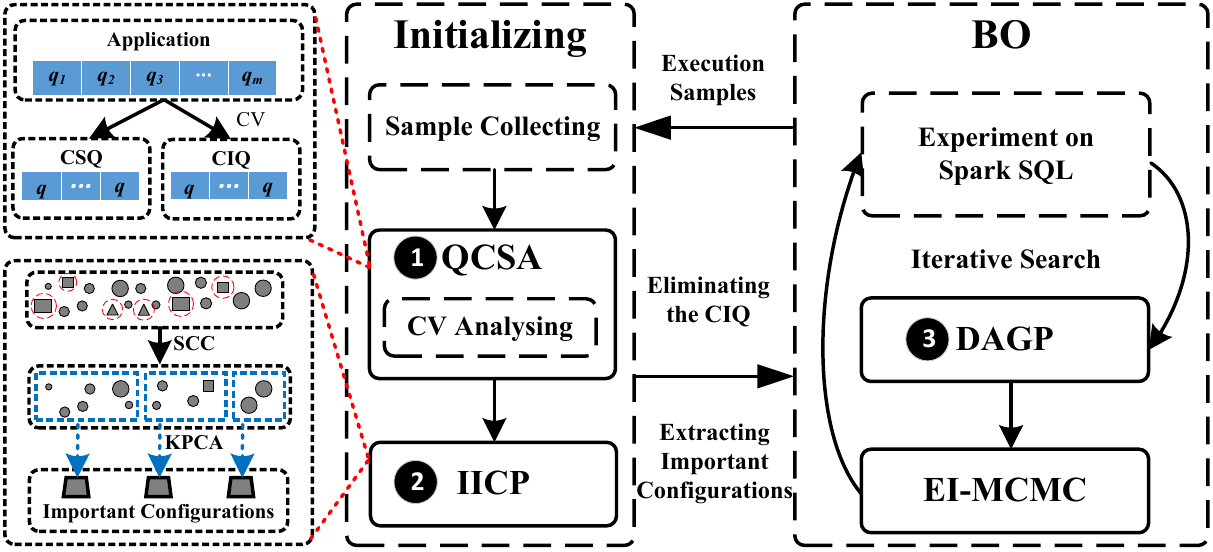}
  \caption{An Overview of LOCAT. BO --- Bayesian Optimization. QCSA --- Query Configuration Sensitivity Analysis. IICP --- Identifying Important Configuration Parameters. DAGP --- Data size Aware Gaussian Process. EI-MCMC --- Expected Improvement with Markov Chain Monte Carlo.}
  \label{fig:locatearch}
  \vspace{-4mm}
\end{figure}

Figure~\ref{fig:locatearch} shows the block diagram of LOCAT. As can be seen, it consists of three components: query configuration sensitive analysis (QCSA), identifying important configuration parameters (IICP), and data-size aware Gaussian Process (DAGP). QCSA analyzes how the performance (e.g., latency) of each query of a Spark SQL application varies when the configuration parameter values change. If the performance of a query varies significantly when parameter values change, we call it configuration sensitive query. Otherwise, we call it configuration insensitive query. IICP identifies the important parameters for a Spark SQL application to be tuned. DAGP models the performance of a Spark SQL application as a Gaussian Process (GP) of the input data size of the application in addition to the configuration parameters.

When we employ LOCAT to optimize the configurations of a Spark SQL application, we firstly leverage QCSA to identify the configuration insensitive queries of the application and in turn remove these queries. We call the resulted application RQA (reduced query application). Subsequently, we use the component IICP to select the important configuration parameters to tune for the RQA. Finally, the selected configuration parameters and the input data size of the RQA are input to the DAGP which is used as the surrogate model of BO to search for the optimal configuration of the RQA. Note that the optimal configuration of the original Spark SQL application is the same as that of the RQA.     

\subsection{Query Configuration Sensitivity Analysis}\label{sec:qcsa}
As aforementioned, ML-based configuration auto-tuning needs to collect a large number of training samples for an application by running the application on a real cluster the same number of times, which is time-consuming. One possible way to reduce the time is to shorten the execution time of each run. Since a Spark SQL application consists of a number of queries, the execution time of the application would be shortened if some queries can be removed from it. The performance of the removed queries should not be influenced by the value variance of the configuration parameters. Moreover, removing queries should not affect the performance of other queries either. However, we do not know which queries of an application can be removed as collecting training samples for it.

\begin{figure}[t!]
  \centering
  \includegraphics[width=\linewidth]{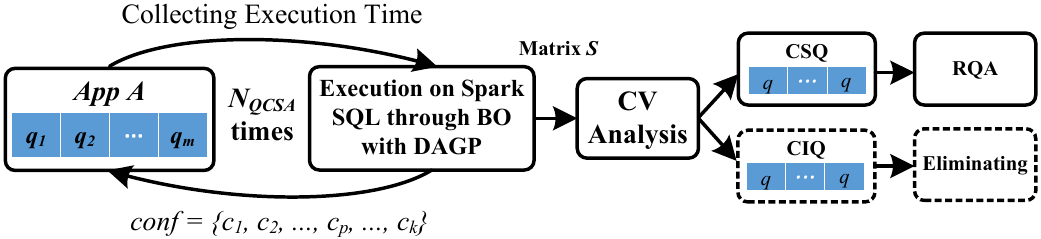}
  \caption{The QCSA diagram. DAGP --- Data size Aware Gaussian Process. CV --- Coefficient of Variation. RQA --- Reduced Query Application. CSQ --- Configuration Sensitive Query. CIQ --- Configuration Insensitive Query.}
  \label{fig:qcsaarch}
  \vspace{-4mm}
\end{figure}

To address this issue, we propose query configuration sensitivity analysis (QCSA) to identify which queries can be removed. Figure~\ref{fig:qcsaarch} shows the block diagram of QCSA. As can be seen, the Spark SQL application $AppA$ executes on a given cluster $N_{QCSA}$ times through BO with DAGP, each with a different configuration. A configuration can be represented by a vector as follows.
\begin{equation}
conf = \{c_1, c_2,...,c_p,...,c_k\}\label{equ:conf}  
\end{equation}
with $c_p$ the $p^{th}$ configuration parameter value and $k$ the total number of configuration parameters. A random configuration is generated by randomly setting the $p^{th}$ value of $conf$ within the $p^{th}$ parameter's value range and $p$ can be any value between 1 and $k$. In $AppA$'s each execution, QCSA records each query's execution time. It is represented by $t_{q_{ij}}$ where $q_{ij}$ denotes the $i^{th}$ query of the $j^{th}$ execution of the $AppA$. After the $AppA$ executes $N_{QCSA}$ times, we have collected a matrix $S$ denoted as follows
\begin{equation}
    S = \{t_{q_{ij}}\}, i=1,2,...,m; j=1,2,...,N_{QCSA}\label{equ:ss}
\end{equation}
with $m$ the number of queries in $AppA$.

We employ Coefficient of Variation (CV), also known as standard deviation divided by mean, to represent the configuration sensitivity of a query $q_i$ because CV is a standard measure of dispersion of a probability distribution or frequency distribution. As such, query $q_i$'s configuration sensitivity is calculated by equation  (\ref{eq:cov}).
\begin{equation}
CV_{q_i}=\frac{1}{\bar{t_{q_i}}}\sqrt{\frac{1}{N_{QCSA}}\sum_{j=1}^{N_{QCSA}}(t_{q_{ij}}-\bar{t_{q_i}})^2}\label{eq:cov}
\end{equation}
with $CV_{q_i}$ the configuration sensitivity of query $q_i$, $\bar{t_{q_i}}$ the average execution time of query $q_i$, $N_{QCSA}$ the total number of executions of the $AppA$ with random configurations, and $t_{q_{ij}}$ the execution time of query $q_i$ in its $j^{th}$ execution. Note that there are $N_{QCSA}$ different $t_{q_i}$ s for $AppA$.

In general, higher $CV_{q_i}$ indicates the corresponding query $q_i$ of a Spark SQL application is more sensitive to configuration tuning. To remove some queries from a Spark SQL application when we collect training samples, we need to determine a suitable threshold of $CV$. However, it is difficult to set a absolute threshold such as $1$ for $CV$ because the value ranges of $CV$ for different queries of the same Spark SQL application might be significantly different, let alone different applications. We therefore need a relative way to determine the threshold for $CV$. It's been proved that classifying $CV$ into {\it high}, {\it medium} and {\it low} is good enough~\cite{lorenzo2015coefficient, vaz2017classification} to leverage $CV$. We therefore equally divide the value range of a $CV$ into three non-overlapped partitions, as shown in equation (\ref{eq:covclass}).
\begin{equation}
Width_{CV}=(max(CV_{q_i}) - min(CV_{q_i}))/3\label{eq:covclass}
\end{equation}
where $Width_{CV}$ is the width of each partition, and $max(CV_{q_i})$ as well as $min(CV_{q_i})$  are the maximum CV and minimum CV of query $q_i$ occurred in the $N_{QCSA}$ executions of application $AppA$, respectively. We classify a query with its $CV\in[0, min(CV_{q_i}) + Width_{CV})$ as a configuration insensitive query (CIQ). Otherwise, the query is a configuration sensitive query (CSQ). To collect training samples for $AppA$, we first remove the CIQs and only remain the CSQs, making $AppA$ the reduced query application (RQA). Subsequently, we run the RQA a number of times, each with a random configuration. With the same configuration, the execution time of RQA is significantly shorter than that of the original $AppA$. As such, we can collect the same number of training samples as that needed to tune the configuration of the original $AppA$ but with dramatically shorter time.  

\subsection{Identifying Important Parameters}\label{sec:siicp}
\begin{figure}[t!]
  \centering
  \includegraphics[width=\linewidth]{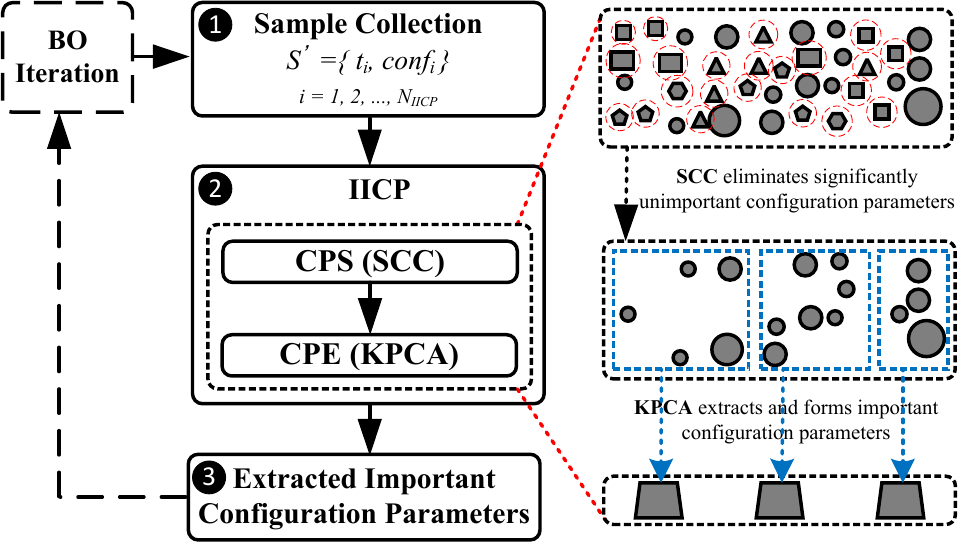}
  \caption{The block diagram of IICP. CPS --- Configuration Parameter Selection. SCC --- Spearman Correlation Coefficient. CPE --- Configuration Parameter Extraction. KPCA --- Kernel Principle Component Analysis.}
  \label{fig:IICPArch}
  \vspace{-4mm}
\end{figure}

As mentioned in Section~\ref{sec:intro}, another way to reduce the optimization time needed by ML-based approaches is to reduce the number of training samples by decreasing the number of parameters needing to be tuned. This is because ML-based approaches typically need to construct highly accurate performance models as functions of parameters. For the same accuracy, more training samples are needed to train a performance model if the model takes more parameters as input. We therefore propose to firstly identify the important configuration parameters (IICP) with respect to performance, and subsequently only select the important ones to build performance models. As such, the number of training samples needed to build high accuracy models can be reduced. Figure~\ref{fig:IICPArch} shows that IICP consists of two stages: sample collection and IICP. 

\subsubsection{The sample collection stage} It collects the execution times of a small number of executions of a $AppA$ with a certain input data size, each execution with a random configuration. The execution times and configurations are stored in a matrix $S'$ shown as:
\begin{equation}
S'=\{t_i, conf_i, ds\}, i=1,2,...,N_{IICP}\label{equ:ssIICP}
\end{equation}
with $t_i$ the execution time of $AppA$ with data size $ds$ executed with $conf_i$ defined by equation (\ref{equ:conf}), and $N_{IICP}$ the number of executions of $AppA$. Smaller $N_{IICP}$ is better because we want to reduce the optimization time.

\subsubsection{The IICP stage}\label{sec:drs} This stage extracts the important configuration parameters in terms of performance based on the samples collected by the sample collection stage. There are a lot of approaches such as metric quantification~\cite{counterminer}, feature selection~\cite{chandrashekar2014survey}, feature extraction~\cite{guyon2008feature} can be used to perform IICP. We do not employ the metric quantification approach used in~\cite{counterminer} because it needs a large number of training samples, which is conflict with our goal. We do not use feature selection and extraction directly in this study either. Instead, we employ a {\it novel hybrid} approach which combines the feature selection and feature extraction.  

We therefore employ two steps: {\it configuration parameter selection} (CPS) and {\it configuration parameter extraction} (CPE) which seem similar but significantly different. CPS removes the unimportant parameters from the vector $conf$ defined by equation (\ref{equ:conf}) and the remaining ones form a new vector shown in equation (\ref{equ:rconf}).
\begin{equation}
    r\_conf = \{c_1,c_2, ..., c_i, ..., c_{rk}\} \label{equ:rconf}
\end{equation}
with $c_i$ the $i^{th}$ configuration parameter and $rk$ the number of remaining parameters after CPS is performed. Note that $rk$ is less than $k$. CPE further extracts important parameters from vector $r\_conf$. Note that these parameters are not the original configuration parameters. Instead, they are new parameters which are functions such as linear regressions of the original ones. These small number of new parameters are used to construct the DAGP of BO in this study. After BO converges, we derive the values of the original configuration parameters from the new parameters to optimally configure a Spark SQL application $AppA$. As such, the time used to search the optimal configuration for $AppA$ can be significantly reduced further.

CPS is implemented by using Spearman Correlation Coefficient (SCC)~\cite{zar2005spearman} which is a popular filter approach for feature selection~\cite{zebari2020comprehensive}. SCC is an efficient multivariate analysis technique without learning involved to measure the strength of association between features~\cite{cheng2021nonlinear}. Compared to Pearson Correlation Coefficient (PCC)~\cite{bolboaca2006pearson}, SCC is more suitable for IICP because the values of configuration parameters tuned in this study are discrete numerical variables. We calculate the SCC between each configuration parameter $c_p$ and the execution time $t_i$, and in turn eliminate $c_p$ if the absolute value of its corresponding SCC is less than $0.2$~\cite{zar2005spearman}, which is a common boundary value of SCC to identify poor correlation~\cite{akoglu2018user,kleinbaum2013applied}. The remaining parameters are stored in $r\_conf$ which is defined by equation (\ref{equ:rconf}). However, the configuration parameters in $r\_conf$ may correlate with each other in a non-linear manner. This indicates that the size of $r\_conf$ can be further reduced but it can not be done by using SCC. We therefore design CPE based on the $r\_conf$ produced by CPS.

Our CPE is performed by Kernel Principal Component Analysis (KPCA) which is a powerful nonlinear feature extractor~\cite{zhang2016big}. The common approach for feature extraction is Principle Component Analysis (PCA). PCA reduces the feature dimension by calculating the eigen vectors of a Covariance matrix of the original inputs. However, PCA can not extract the non-linear information from the original configuration space~\cite{mika1998kernel}. KPCA extends PCA to make it be able to extract non-linear information by leveraging the kernel method. The crucial problem of KPCA is to select a suitable kernel and we select it by experiments. If we use the configuration parameters selected by KPCA with different kernels to configure a Spark SQL application to execute a number of times, the larger standard deviation (SD) of the execution times caused by a kernel indicates that the configuration parameters selected by the kernel are more important to execution time than those selected by other kernels. 

\begin{figure}[t!]
  \centering
  \includegraphics[width=2.5in]{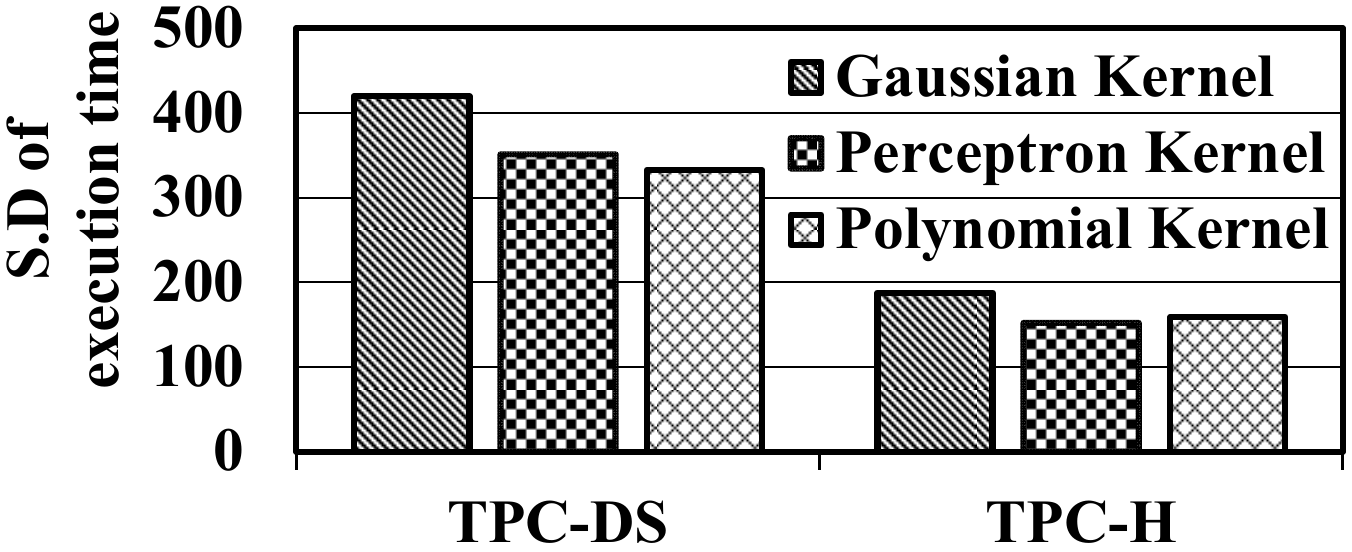}
  \vspace{-4mm}
  \caption{Kernel comparison. S.D --- standard deviation.}
  \label{fig:knlslt}
  \vspace{-4mm}
\end{figure}

We evaluate three mainstream kernel methods: Gaussian kernel, perceptron kernel, and polynomial kernel~\cite{hao2013learning} in our experimental environment (Section~\ref{sec:exp}) for two Spark SQL applications: $TPC-DS$ and $TPC-H$. As shown in Figure~\ref{fig:knlslt}, the SDs of execution times caused by the Gaussian kernel are the largest for both $TPC-DS$ and $TPC-H$. This indicates that the configuration parameters selected by KPCA with the Gaussian kernel are more important than those selected with other kernels in terms of performance. We therefore choose Gaussian kernel in this study.  

\subsection{Datasize-Aware Bayesian Optimization}
Previous BO-based configuration optimization approaches such as CherryPick~\cite{alipourfard2017cherrypick} can not adapt to the input data size changes of an application. These approaches therefore can not be used online. To address this issue, we design Datasize-Aware Gaussian Process (DAGP) for BO to make it be able to adapt to data size changes.

In detail, we employ GP to model the execution time of a Spark SQL application with a certain configuration as a function of configuration parameters and input data size as follows.
\begin{equation}
    t = f(conf,ds)
\end{equation}
with $t$ the execution time of a Spark SQL application, $conf$ the configuration vector defined by equation (\ref{equ:conf}) used to configure the application, and $ds$ the input data size. As such, LOCAT can adapt to the data size changes of a Spark SQL application during optimization. We express the vector $\{conf,ds\}$ as $X_e$.

We now elaborate the function $f$ by Gaussian distribution~\cite{rasmussen2003gaussian} as equation~(\ref{eq:0GP}).
\begin{equation}
f(X_e)\sim GP(0,K(X_E,X'_E))
\label{eq:0GP}
\end{equation}
where $K$, the kernel function, denotes a covariance matrix. After we run a Spark SQL application with $n$ configurations with different input data sizes, we get a matrix $(X_E, T)$ with $n$ $X_e$ and corresponding $t$ as input training set of GP. The $(X_E, T)$ can be expressed by ($\{X_{e1},X_{e2},...,X_{en}\}$,$\{t_1,t_2,...,t_n\}$). $X'_E$ and $X_E$ are two matrices of the same size. We solve the regression model posterior under Gaussian process prior and normal likelihood: $p[f(X_e)|f(X_{e1}),f(X_{e2}), ...,$ $f(X_{en}))]$ to get the joint probability distribution of $t$, the actual output value of the training sample, and $f_*$, the predicted value of a sample, as equation~(\ref{eq:unitedDis}).

\begin{equation}\label{eq:unitedDis}
            \begin{bmatrix}
               t  \\
               f_*
             \end{bmatrix}
            =N
            \begin{bmatrix}
            0,
            \begin{bmatrix}
               K(X_E,X_E)+{{\delta_n}^2}I & K(X_E,{X_E}_*)  \\
               K(X_E,X_E) & K({X_E}_*,{X_E}_*)
             \end{bmatrix}\\
            
            \end{bmatrix}
\end{equation}
where ${X_E}_*$ is a set of samples to be predicted. $I$ is a $n$ dimensional identity matrix. ${\delta_n}^2$ is the variance of noise. 

Next, we take the edge distribution of $f_*$ for the joint probability distribution expressed by equation (\ref{eq:unitedDis}), and the regression prediction form of the DAGP can be obtained from the edge distribution property of the joint normal distribution as equation~(\ref{eq:unitPre}):

\begin{equation}\label{eq:unitPre}
\left\{
             \begin{array}{lr}
             p(f_*|X_E,t,{X_E}_*,{{\delta_n}^2})=N[f_*|\bar{f_*},cov(f_*)], &  \\
             \bar{f_*}=K({X_E}_*,X_E)(K(X_E,X_E)+{{\delta_n}^2}I)^{-1}t, &  \\
             cov(f_*)=K({X_E}_*,{X_E}_*)-K({X_E}_*,X_E)\\ \ \ \ \  \ \ \ \ \ \ \  (K(X_E,X_E)+{{\delta_n}^2}I)^{-1}K({X_E}_*,X_E), &  
             \end{array}
\right.
\end{equation}

\textbf{Acquisition function:} We use the Expected Improvement (EI) with Markov Chain Monte Carlo (MCMC) hyperparameter marginalization algorithm~\cite{snoek2012practical} as BO's acquisition function, which shows better performance than other acquisition functions across wide test cases~\cite{snoek2012practical}. BO uses the EI-MCMC to avoid external tuning of GP's hyperparameters, and iteratively selects the next configuration sample with the greatest potential to minimize the execution time of a Spark SQL application. 

\textbf{Start points:} LOCAT incrementally builds the GP model, starting with three samples generated by Latin Hypercube Sampling (LHS)~\cite{helton2003latin}. After each execution, the GP model is improved and helps BO pick the next candidate configuration that is estimated to minimize the execution time of a Spark SQL application.

\textbf{Stop condition:} The GP modeling stops after at least 10 iterations and the EI drops below $10\%$. The goal of setting stop condition is to balance between the exploration of configuration space $X_e$ and the exploitation around the optimal configuration found thus far, which is inspired by CherryPick~\cite{alipourfard2017cherrypick}.

In summary, BO starts with the training samples selected by LHS and employs the samples to initialize DAGP. BO then continuously takes more samples recommended by the DAGP with EI-MCMC until the stop condition is met. QCSA and IICP are designed to accelerate the optimization process of BO.

\section{Experimental Setup}\label{sec:exp}

\subsection{Experimental Clusters and Framework}
To evaluate LOCAT, we employ two significantly different clusters: an ARM cluster and an x86 cluster. The ARM cluster consists of four KUNPENG ARM servers. One serves as the master node and the other three servers serve as slave nodes. Each server is equipped with 4 KUNPENG 920 2.60GHz 32-core processors and 512GB PC4 memory. There are in total 512 cores and 2,048 GB memory in the ARM cluster. The x86 cluster consists of eight Xeon severs and one server serves as the master node and the other seven servers are slave nodes. Each x86 server has 2 Intel(R) Xeon(R) Silver 4114 2.20GHz ten-core processors and 64GB PC4 memory. There are in total 160 cores and 512 GB memory in the x86 cluster. The choice of an ARM cluster and an x86 cluster is to evaluate how well LOCAT can adapt to different hardware. Using a four-node and an eight-node cluster is to validate how well LOCAT works in different scales of clusters. On the two clusters, we use Spark 2.4.5 as our experimental framework because of Spark 2.4.5 is more steady and popular in industry compared to other versions. 

\subsection{Representative Programs}
We select the $TPC-DS$~\cite{tpcds}, $TPC-H$~\cite{boncz2013tpc}, and three programs from $HiBench$~\cite{hibench} as representative programs to evaluate LOCAT, as shown in Table~\ref{tbl:benchmarks}. $TPC-DS$, containing $104$ queries, has been widely used in Spark SQL systems for research and development of optimization techniques~\cite{chiba2018towards,ivanov2015evaluating,ramdane2018partitioning}. It models complex decision support functions to provide highly comparable, controlled, and repeatable tasks in evaluating the performance of Spark SQL systems~\cite{barata2015overview}.  
$TPC-H$ benchmark is similar to $TPC-DS$ that simulates a decision support system database environment. We select the $TPC-H$ because it can represent a near-real analysis business with 22 queries only, which is less than $TPC-DS$.

The $HiBench$ benchmark suite has been widely used to evaluate the Spark framework and we select three SQL related benchmarks with a single query each in this study: Join, Scan, and Aggregation. 1) Join is a query that typically executes in two phases: Map and Reduce. 2) Scan is a query that consists of only Map operation initiated by the "select" command that splits the input value based on the field delimiter and outputs a record. 3) Aggregation is a query that consists of both Map and Reduce operations. The Map operation ("select" command) first splits the input value by the field delimiter and then outputs the field defined by the Reduce operation("group by" command) as a new key/value pair. In our experiment, we treat these three workloads as three separate benchmarks named {\it Join}, {\it Scan}, and {\it Aggregation}. To evaluate how LOCAT adapts to the dynamic changes of input data size, we employ five different data sizes for our experiments (100GB, 200GB, 300GB, 400GB, and 500GB). 

\begin{table}[t!]
\caption{Experimented Benchmarks and Input Data Sizes.}
  \label{tab:freq}
  \vspace{-2mm}
  \begin{tabular}{|l|c|}
    \hline
    Benchmark&Input Data Size\\
    \hline
    TPC-DS & \multirow{5}{*}{100, 200, 300, 400, 500 (GB)}\\
    \cline{1-1}
    TPC-H&\\
    \cline{1-1}
    HiBench Join &  \\
    \cline{1-1}
    HiBench Scan &\\
    \cline{1-1}
    HiBench Aggregation &\\
    \hline
\end{tabular}\label{tbl:benchmarks}
\vspace{-4mm}
\end{table}

\begin{table*}\small
  \caption{Description of Selected Parameters.}
  \label{table:parameters}
  \setlength{\tabcolsep}{1mm}
  \begin{tabular}{|l|c|c|c|}
    \hline
    Configuration Parameters--Description & Default & Range A & Range B\\
    \hline
    \textbf{spark.broadcast.blockSize} -- Specifies the size of each piece of a block for TorrentBroadcastFactory, in MB. & 4& 1 - 16 & 1 - 16 \\
    \hline
    \textbf{spark.default.parallelism} -- Specifies the maximum number of partitions in a parent RDD for shuffle operations. & \#& 100 - 1000 & 100 - 1000 \\
    \hline
    \textbf{*spark.driver.cores} -- Specifies the number of cores to use for the driver process. & 1& 1 - 8 & 1 - 16\\
    \hline
    \textbf{*spark.driver.memory} -- Specifies the amount of memory to use for the driver process, in GB. & 1 & 4 - 32 & 4 - 48\\
    \hline
    \textbf{*spark.executor.cores} -- Specifies how many CPU cores each executor process uses. & 1 & 1 - 8 & 1 - 16 \\
    \hline
    \textbf{spark.executor.instances} -- Specifies the total number of Executor processes used for the Spark job. & 2 & 48 - 384 & 9 - 112 \\
    \hline
    \textbf{*spark.executor.memory} -- Specifies how much memory each executor process uses, in GB. & 1 & 4 -32 & 4 - 48\\
    \hline
    \textbf{*spark.executor.memoryOverhead} -- Specifies the additional memory size to be allocated per executor, in MB. & 384 & 0 - 32768 & 0 - 49152\\
    \hline
    \textbf{spark.io.compression.zstd.bufferSize} -- Specifies the buffer size used in Zstd compression, in KB. & 32 & 16 - 96 & 16 -96 \\
    \hline
    \textbf{spark.io.compression.zstd.level} -- Specifies the compression level for Zstd compression codec. & 1 & 1 -5 & 1 - 5\\
    \hline
    \textbf{spark.kryoserializer.buffer} -- Specifies the initial size of Kryo's serialization buffer, in KB. & 64 & 32 - 128 & 32 - 128\\
    \hline
    \textbf{spark.kryoserializer.buffer.max}-- Specifies the maximum allowable size of Kryo serialization buffer, in MB. & 64 & 32 -128 & 32 - 128\\
    \hline
    \textbf{spark.locality.wait}-- Specifies the wait time to launch a task in a data-local before in a less-local node, in second. & 3 & 1 - 6 & 1 - 6 \\
    \hline
    \textbf{spark.memory.fraction} -- Specifies the fraction of (heap space - 300MB) used for execution and storage. & 0.6 & 0.5 - 0.9 & 0.5 - 0.9\\
    \hline
    \textbf{spark.memory.storageFraction} -- Specifies the amount of storage memory immune to eviction. & 0.5 & 0.5 - 0.9 & 0.5 - 0.9\\
    \hline
    \textbf{*spark.memory.offHeap.size} -- Specifies the memory size\textbf{} which can be used for off-heap allocation, in MB. & 0 & 0 - 32768 & 0 - 49152\\
    \hline
    \textbf{spark.reducer.maxSizeInFlight} -- Specifies the maximum size to fetch simultaneously from a reduce task, in MB. & 48 & 24 - 144 & 24 - 144 \\
    \hline
    \textbf{spark.scheduler.revive.interval} -- Specifies the interval for the scheduler to revive the worker resource, in second. & 1 & 1 - 5 & 1 - 5\\
    \hline
    \textbf{spark.shuffle.file.buffer} -- Specifies in-memory buffer size for each shuffle file output stream, in KB. & 32 & 16 - 96 & 16 -96\\
    \hline
    \textbf{spark.shuffle.io.numConnectionsPerPeer} -- Specifies the amount of connections between hosts are reused. & 1 & 1 -5 & 1 - 5\\
    \hline
    \textbf{spark.shuffle.sort.bypassMergeThreshold} -- Specifies the partition number to skip mapper side sorts. & 200 & 100 - 400 & 100 - 400\\
    \hline
    \textbf{spark.sql.autoBroadcastJoinThreshold} -- Specifies the maximum size for a broadcasted table, in KB. & 1024 & 1024 - 8192 & 1024 - 8192 \\
    \hline
    \textbf{spark.sql.cartesianProductExec.buffer.in.memory.threshold} -- Specifies row numbers of Cartesian cache. & 4096 & 1024 - 8192 & 1024 - 8192\\
    \hline
    \textbf{spark.sql.codegen.maxFields} -- Specifies the maximum field supported before activating the entire stage codegen. & 100 & 50 - 200 & 50 - 200\\
    \hline
    \textbf{spark.sql.inMemoryColumnarStorage.batchSize} -- Specifies the size of the batch used for column caching. & 10000 & 5000 - 20000 & 5000 - 20000\\
    \hline
    \textbf{spark.sql.shuffle.partitions} -- Specifies the default partition number when shuffling data for joins or aggregations. & 200 & 100 - 1000 & 100 - 1000\\
    \hline
    \textbf{spark.storage.memoryMapThreshold} -- Specifies mapped memory size when read a block from the disk, in MB. & 1 & 1 - 10 & 1 - 10\\
    \hline
    spark.broadcast.compress -- Decides whether to compress broadcast variables before sending them. & true & true, false & true, false\\
    \hline
    spark.memory.offHeap.enabled -- Decides whether to use off-heap memory for certain operations. & true & true, false & true, false\\
    \hline
    spark.rdd.compress -- Decides whether to compress serialized RDD partitions. & true & true, false & true, false\\
    \hline
    spark.shuffle.compress -- Decides whether to compress map output files. & true & true, false & true, false\\
    \hline
    spark.shuffle.spill.compress -- Decides whether to compress data spilled during shuffles. & true & true, false & true, false\\
    \hline
    spark.sql.codegen.aggregate.map.twolevel.enable -- Decides whether to enable two-level aggregate hash mapping. & true & true, false & true, false\\
    \hline
    spark.sql.inMemoryColumnarStorage.compressed -- Decides whether to compress each column based on data. & true & true, false & true, false\\
    \hline
    spark.sql.inMemoryColumnarStorage.partitionPruning -- Decides whether to prune partition in memory. & true & true, false & true, false\\
    \hline
    spark.sql.join.preferSortMergeJoin -- Decides whether to use sort Merge Join instead of Shuffle Hash Join. & true & true, false & true, false\\
    \hline
    spark.sql.retainGroupColumns -- Decides whether to retain group columns. & true & true, false & true, false\\
    \hline
    spark.sql.sort.enableRadixSort -- Decides whether to use radix sort. & true & true, false & true, false\\
    \hline
  \end{tabular}
\end{table*}

\subsection{Configuration Parameters}
Table~\ref{table:parameters} shows the configuration parameters of Spark SQL applications we considered in this study. The first column of Table~\ref{table:parameters} describes the parameter names and their definitions. The second column provides the default value of each configuration parameter which is recommended by the Spark official website~\cite{SparkConfWebsite}. The columns "Range A" and "Range B" show the value ranges of each configuration parameter on the ARM cluster and the x86 cluster, respectively. The value range determination for each parameter is presented in Section~\ref{sec:vrd}.

\section{Results and Analysis}\label{sec:rlt}
In this section, we first determine the numbers of initial experimental samples, $N_{QCSA}$ and $N_{IICP}$, needed by LOCAT. Subsequently, we present the results and analysis.  

\begin{figure}
  \centering
  \includegraphics[width=3in]{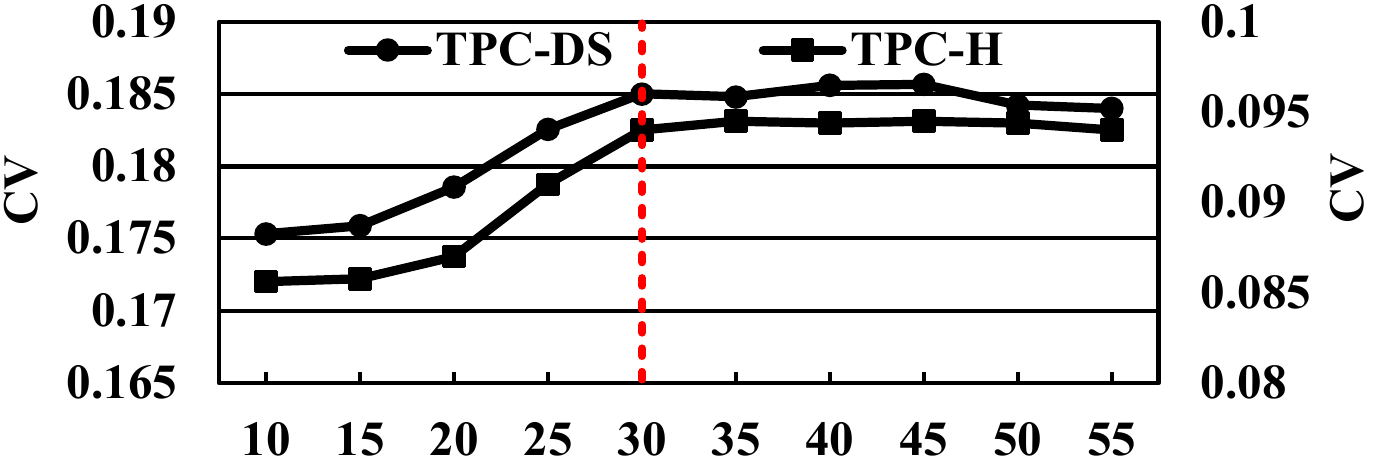}
  \vspace{-2mm}
  \caption{How CV (Coefficient of Variation) changes along with the increasing number of experimental samples for QCSA. The left and right Y axes represent the CVs of $TPC-DS$ and $TPC-H$, respectively. }
  \label{fig:QCSAsampleScale}
  \vspace{-4mm}
\end{figure}

\begin{figure}[t!]
  \centering
  \includegraphics[width=\linewidth]{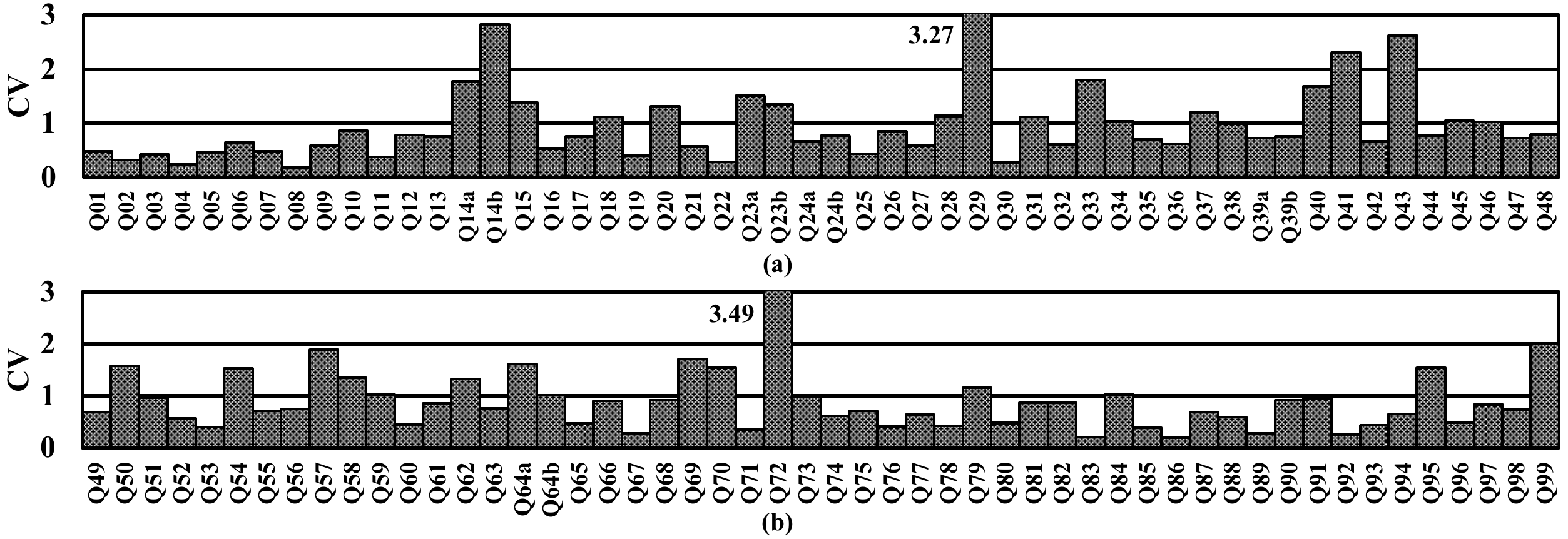}
  \vspace{-6mm}
  \caption{Configuration Sensitivity denoted by CV (coefficient variation) of the $TPC-DS$ queries. Y axis denotes the CVs of queries when configurations are changed.}\label{fig:querySen}
  \vspace{-6mm}
\end{figure}

\subsection{Determining $N_{QCSA}$}
As mentioned in Section~\ref{sec:qcsa}, we need $N_{QCSA}$ of experimental samples to perform the query configuration sensitivity analysis (QCSA). To make the time used to collect experimental samples as short as possible, $N_{QCSA}$ should be as small as possible while it should also be large enough to accurately reflect the CV of the Spark SQL queries. We employ experiments to determine a suitable value of $N_{QCSA}$ to satisfy the above requirements. Figure~\ref{fig:QCSAsampleScale} shows that how the CVs for $TPC-DS$ and $TPC-H$ change when we increase the number of experimental samples. As can be seen, when $N_{QCSA}$ increases from 10 to 30, the CV for $TPC-DS$ as well as that for $TPC-H$ keep increasing. When $N_{QCSA}$ is larger than 30, the CVs for both $TPC-DS$ and $TPC-H$ do not increase any longer. This indicates that 30 samples are enough for QCSA and more samples do not provide any information for CV besides wasting time. We therefore set $N_{QCSA}$ to 30 in this study. Note that we do not collect additional 30 experimental samples before we perform BO with DAGP (Dataisize-Aware Gaussian Process). Instead, we leverage the samples (executions) performed by the BO iterations. 

\subsection{QCSA Results and Analysis}
After we set $N_{QCSA}$ to 30, we perform QCSA for $TPC-DS$ based on the 30 experimental samples. Figure~\ref{fig:querySen} shows the CVs for the $104$ queries of $TPC-DS$. A couple of interesting findings can be observed here. For one, {\it the CVs for different queries are significantly different}. For example, the CV of query {\tt Q04} is only 0.24 while that of query {\tt Q72} is 3.49. We call the queries with small CVs configuration insensitive queries (CIQ) while others configuration sensitive queries (CSQ). This indicates that the performance of CIQs such as {\tt Q04} does not change much when the configuration changes while that of CSQs such as {\tt Q72} does. Second, {\it long queries are not necessarily sensitive to configuration tuning}. For example, the CV of query {\tt Q04} is relatively small (0.24) and its execution time is relatively long (e.g.,80 seconds) while the CV of query {\tt Q14b} is relatively large (2.8) and its execution time is also relatively long (e.g.,49 seconds). This implies that removing long CIQs such as {\tt Q04} can significantly reduce the sample collection time when we collect experimental samples. This also indicates that tuning the configuration of long CSQs such as {\tt Q14b} can improve performance more than tuning short queries with similar CVs. Why are some queries sensitive to configuration tuning while others are not? The reason is analyzed in Section~\ref{sec:rcsqciq}.

Based on these findings, we remove queries by using the CV-based criteria introduced in Section~\ref{sec:qcsa} for experimental sample collection. For the $104$ queries in $TPC-DS$, we remove 81 queries and remain 23 queries when we collect experimental samples. The remaining 23 queries are {\tt \{Q72, Q29, Q14b, Q43, Q41, Q99, Q57, Q33, Q14a, Q69, Q40, Q64a, Q50, Q21, Q70, Q95, Q54, Q23a, Q23b, Q15, Q58, Q62, Q20\}}. That is, we only execute 23 queries in each BO iteration with a different configuration during we search the optimal configuration for $TPC-DS$. As such, the time used to collect the experimental samples can be significantly reduced.

\subsection{Determining $N_{IICP}$}
To perform IICP, we need $N_{IICP}$ of experimental samples to observe how the performance of a Spark SQL application changes according to the value changes of each configuration parameter. Like $N_{QCSA}$, the value of $N_{IICP}$ should be as small as possible because our goal is to make the time for collection experimental samples short. On the other hand, $N_{IICP}$ should also be large enough to correctly identify the important parameters with respect to performance. 

\begin{figure}[t!]
 \centering
 \includegraphics[width=2.8in]{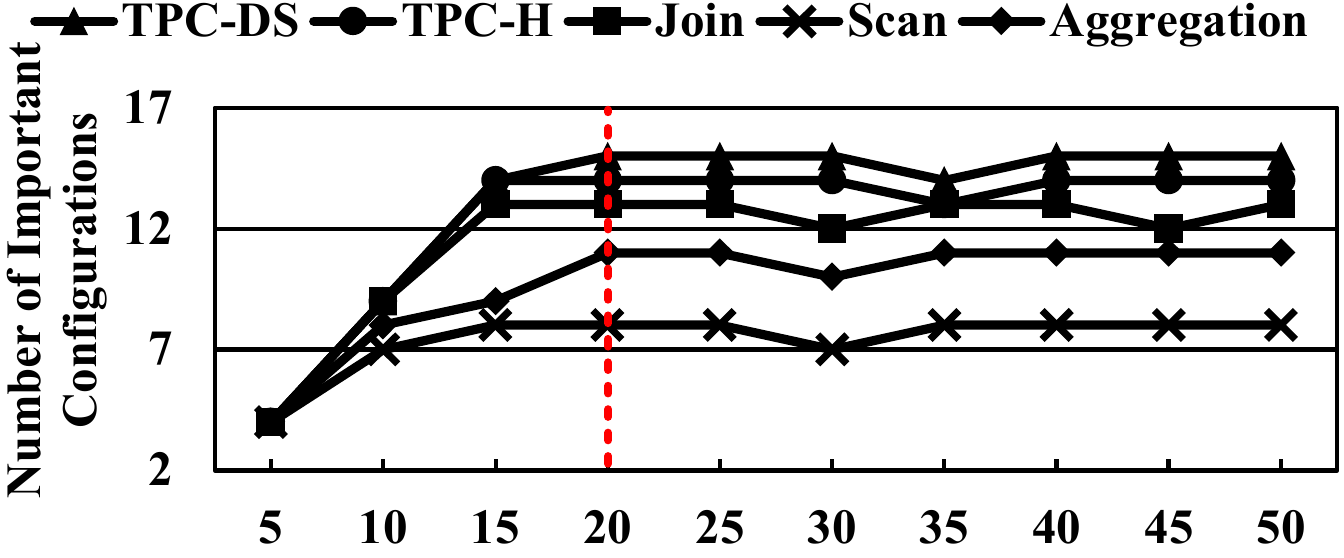}
  \vspace{-4mm}
 \caption{The number variation of identified important parameters along with the increasing number of samples.}
 \label{fig:n_iicp}
 \vspace{-4mm}
\end{figure}

Again, we employ experiments to determine a suitable value for $N_{IICP}$. At the first step, we set $N_{IICP}$ to 5 and we therefore run a Spark SQL application five times, each time with a random configuration. The execution times of the five executions and their corresponding configurations are stored in matrix $S'$ defined by equation (\ref{equ:ssIICP}). We then leverage CPS and CPE described in Section~\ref{sec:drs} to identify the important configuration parameters with respect to performance. We repeat this step a number of times with each time increasing $N_{IICP}$ by 5. We subsequently observe the number of the identified important configuration parameters. If the number of the parameters keeps constant and parameters remain the same when we perform the IICP with increasing values of $N_{IICP}$, it indicates that larger $N_{IICP}$ does not help correctly identify important configuration parameters. In our experiments, we tried ten values of $N_{IICP}$ (5, 10, 15, 20, 25, 30, 35, 40, 45, and 50). Figure~\ref{fig:n_iicp} shows that the number of the identified important configuration parameters for $TPC-DS$ keeps the same when the value of $N_{IICP}$ is equal to or larger than 
20. In addition, the important parameters are also the same when $N_{IICP}$ is larger than 20. We also perform the same experiment for $TPC-H$, Join, Scan, and Aggregation. We find the same phenomenon as shown in Figure~\ref{fig:n_iicp} and we therefore set $N_{IICP}$ to 20 which is less than the value 30 that we set for $N_{QCSA}$. Note that we also take the executions in BO iterations as the experimental samples to perform IICP. Therefore, 30 experimental samples generated by BO iterations are enough for performing the QCSA and IICP of LOCAT.    

\begin{figure}[t!]
  \centering
  \includegraphics[width=3in]{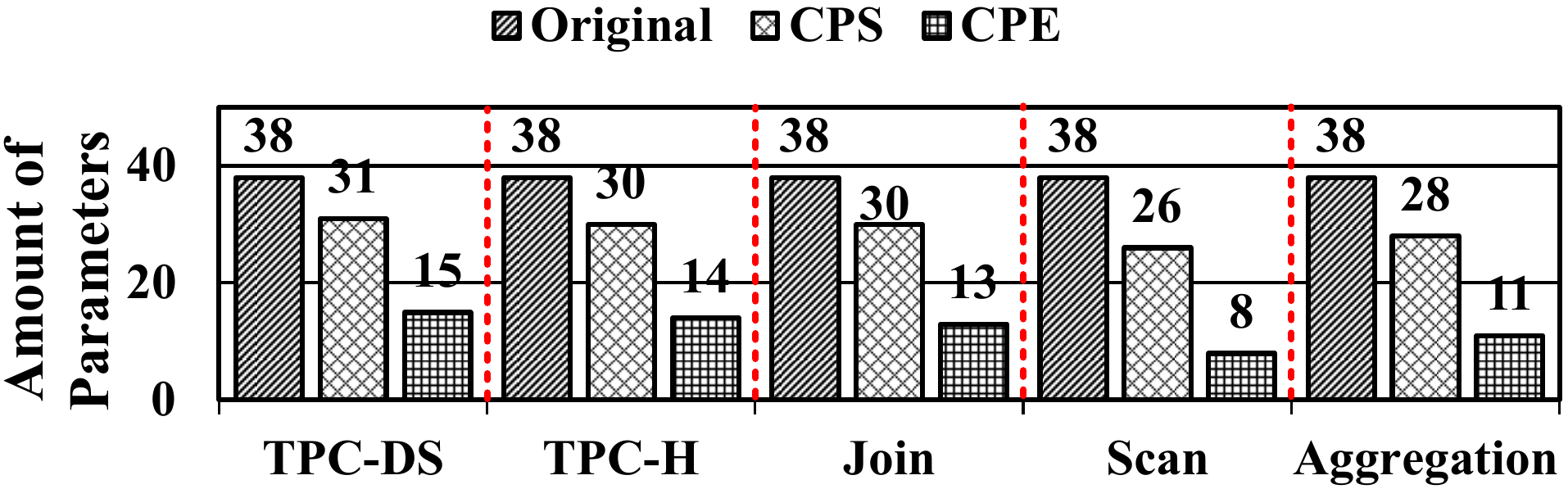}
  \vspace{-4mm}
  \caption{The number of important configuration parameters selected by CPS and CPE.}
  \label{fig:cps_cpe}
  \vspace{-4mm}
\end{figure}

Figure~\ref{fig:cps_cpe} shows the number of important configuration parameters identified by CPS and the ones further extracted by CPE. As can be seen, CPS selects about 2/3 of the original 38 configuration parameters as the important configuration parameters for five Spark SQL applications. CPE further extracts about 1/3 of the important configuration parameters selected by CPS. As a result, the number of configuration parameters fed to GP is significantly reduced and in turn the time used to search for the optimal configuration is accordingly dramatically decreased. 

\begin{table}[t!]\small
  \caption{Top 5 important configurations selected by $CPS$ with $100GB$, $500GB$, and $1TB$ input data size of $TPC-DS$.}
  \label{tab:freq}
  \vspace{-4mm}
  \setlength{\tabcolsep}{0.5mm}
  \begin{tabular}{|c|c|c|c|}
    \hline
    Datasize&100GB&500GB&1TB\\
    \hline
    &sql.shuffle.partitions&sql.shuffle.partitions&sql.shuffle.partitions\\
    \cline{2-4}
    Conf&executor.memory&shuffle.compress&shuffle.compress\\
    \cline{2-4}
    (spark.)&executor.cores&executor.memory&executor.memory\\
    \cline{2-4}
    &shuffle.compress&executor.instances&executor.instances\\
    \cline{2-4}
    &executor.instances&executor.cores&memory.offHeap.size\\
    \hline
\end{tabular}
\label{tbl:importantConfs}
\vspace{-2mm}
\end{table}

\begin{figure}[t!]
  \centering
  \includegraphics[width=\linewidth]{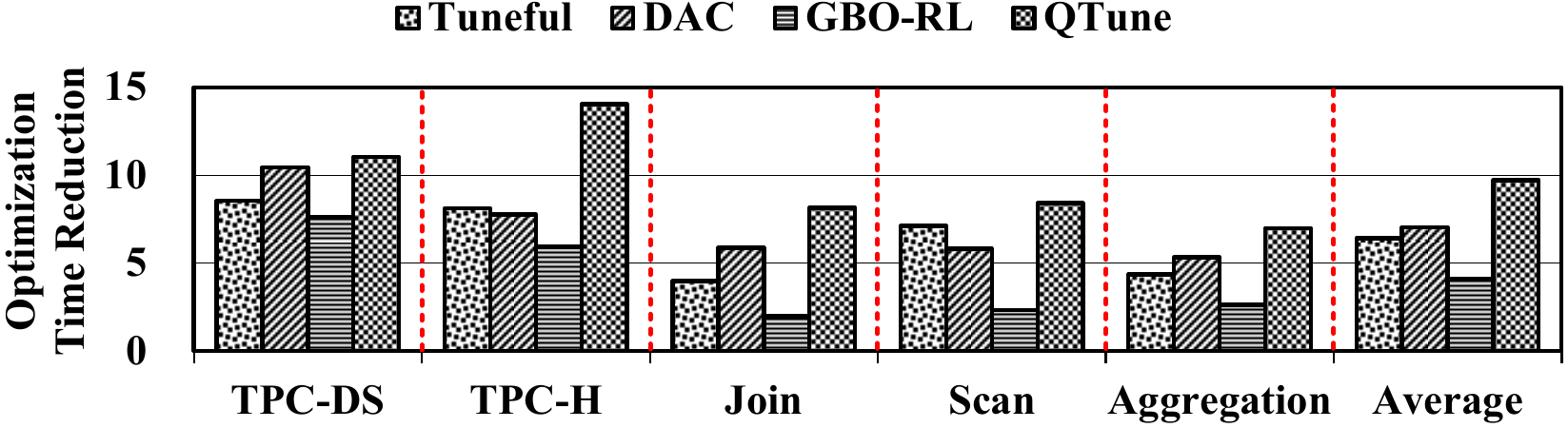}
\vspace{-6mm}
  \caption{Optimization time comparison between LOCAT and others on the four-node ARM cluster. Y axis denotes the time reduction which is defined by using the optimization time taken by LOCAT to divide those taken by others.}
  \label{fig:overhead}
  \vspace{-4mm}
\end{figure}

\begin{figure}[t!]
  \centering
  \includegraphics[width=\linewidth]{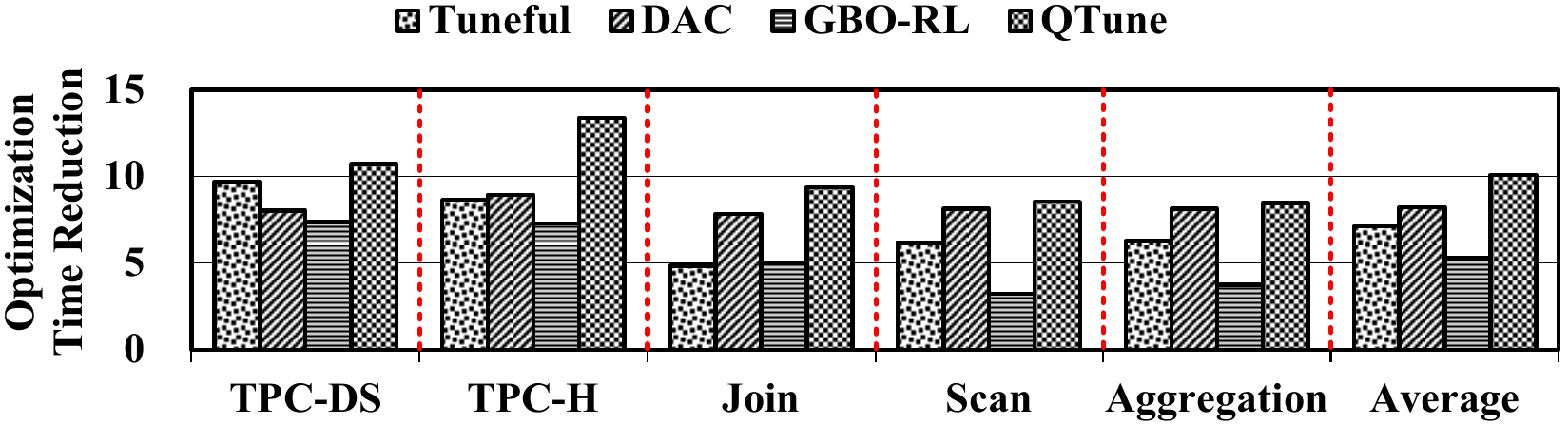}
  \vspace{-6mm}
  \caption{Optimization Time Comparison between LOCAT and others on the eight-node x86 cluster. Y axis denotes the time reduction which is defined by using the optimization time taken by LOCAT to divide those taken by others.}
  \label{fig:overheadCom-8nodes}
  \vspace{-6mm}
\end{figure}

\subsection{Important Parameter Examples}
By using the technique CPS described in Section~\ref{sec:siicp}, we identify 15 important configuration parameters for the experimented benchmarks. Due to the space limitation, we show the five most important parameters for $TPC-DS$ with three input data sizes in Table~\ref{tbl:importantConfs}. A couple of interesting findings can be made here. For one, the most important parameters for the three significantly different input data sizes are all {\it spark.sql.shuffle.partitions}. This parameter specifies the default number of partitions to use when shuffling data for joins or aggregations. Theoretically, this parameter's value significantly influences the parallelism of shuffle operations, which in turn dramatically impacts the performance of a Spark SQL application.

Second, the three parameters related to the number of executor instances, memory size, and whether compress should be applied on shuffle operations are always in the top five important ones for the three input data sizes, their orders might be different though. The number of executor instances influences the task parallelism; the memory size controls the amount of memory can be used by Spark SQL tasks; and the compress influences the amount of data moved between the servers in the cluster, as well as between the memory and disks. Naturally, these aspects influence the performance of a Spark SQL application significantly.

Last but not the least, the parameter {\it Spark.memory.offHeap.size} comes to the fifth most important parameter when the data size increases to $1TB$. This is reasonable because the amount of memory can be used for off-heap allocation becomes important for performance when the input data size is large enough (e.g., $1TB$). In summary, these findings indicate that the important parameters with respect to performance found by IICP are convincible because we can find reasonable explanations for them.

\subsection{Optimization Time}
Figure~\ref{fig:overhead} shows the optimization time reduction achieved by LOCAT on the ARM cluster, which is defined by using the optimization time taken by LOCAT to divide those taken by Tuneful, DAC, GBO-RL, and QTune. Note that the input data sizes for the benchmarks are all 300GB. As can be seen, the time taken by LOCAT to achieve the optimal performance of all benchmarks is significantly shorter than those used by other approaches. In detail, the times taken by Tuneful, DAC, GBO-RL, and QTune are $6.4\times$, $7.0\times$, $4.1\times$, and $9.7\times$ of the time used by LOCAT on average, and up to $7.9\times$, $8.9\times$, $6.3\times$, and $11.8\times$, respectively. Figure~\ref{fig:overheadCom-8nodes} shows the results on the x86 cluster. As can be seen, LOCAT reduces the optimization time taken by Tuneful, DAC, GBO-RL, and QTune by factors of $6.4\times$, $6.3\times$, $4.0\times$, and $9.2\times$ on average and up to $9.7\times$, $8.0\times$, $7.0\times$, and $10.3\times$, respectively. These results indicate two insights. First, LOCAT can indeed significantly reduce the time used by ML approaches to optimize the performance of a wide range of Spark SQL applications. Second, LOCAT can adapt to significantly different hardware as well as different scale of clusters.

The optimization time reduction made by LOCAT comes from LOCAT's three innovations. 1) It leverages QCSA to eliminate the executions of configuration-insensitive queries in BO iterations, which significantly reduces the time for executing a Spark SQL application in each BO iteration. As a result, LOCAT significantly reduces the time used for collecting experimental samples. 2) LOCAT accelerates the BO convergence by developing IICP to reduce the dimension of the configuration searching space. 3) LOCAT leverages DAGP to adapt to the data size changes in optimization process, which enables LOCAT to reuse prior results with different input data sizes and in turn reduce the time overhead.

\begin{figure*}[t!]
  \centering
  \includegraphics[width=\linewidth]{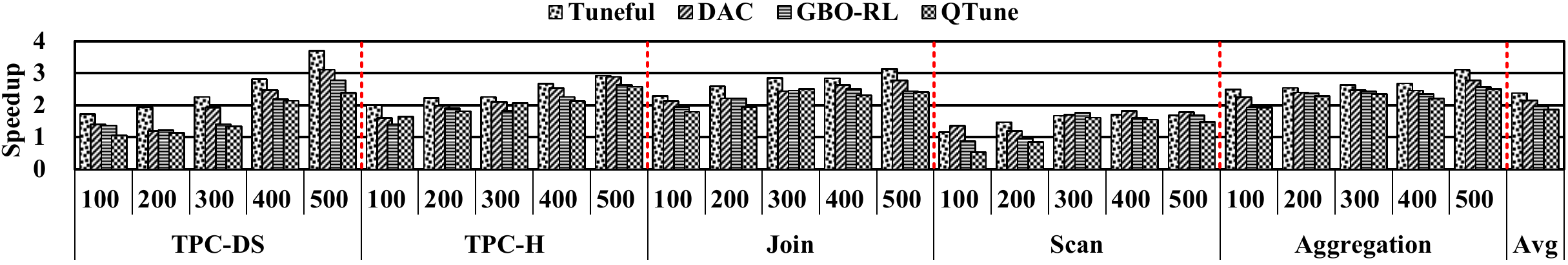}
  \vspace{-6mm}
  \caption{Speedups of the performance tuned by LOCAT over those tuned by Tuneful, DAC, GBO-RL, and QTune on the four-node ARM cluster. The unit of the numbers along with the X axis is GB. The Y axis represents the speedup which is defined by using the execution time of a program-input pair tuned by LOCAT to divide that of it tuned by another approach.}
  \label{fig:speedup}
  \vspace{-2mm}
\end{figure*}

\begin{figure*}[t!]
  \centering
  \includegraphics[width=\linewidth]{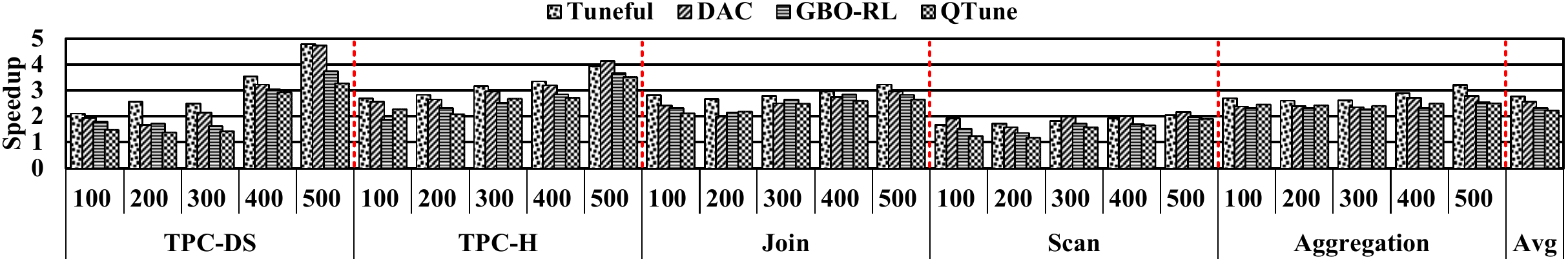}
  \vspace{-6mm}
  \caption{Speedups of the performance tuned by LOCAT over those tuned by other approaches on the eight-node x86 cluster. The unit of the numbers along with the X axis is GB. The Y axis represents the speedup which is defined by using the execution time of a program-input pair tuned by LOCAT to divide that of it tuned by another approach.}
  \label{fig:speedup-8nodes}
  \vspace{-2mm}
\end{figure*}

\subsection{Speedup}\label{sec:speedup}
Although LOCAT significantly reduces the optimization time needed by the state-of-the-art (SOTA) approaches, it is still unclear if it can achieve the performance tuned by the SOTA approaches. In this section, we compare the speedups of the program-input pairs tuned by LOCAT over they tuned by Tuneful, DAC, GBO-RL, and QTune. The speedup is defined as
\begin{equation}
    speedup = \frac{ET_{sota}}{ET_{locat}}
\end{equation}\label{equ:speedup}
with $ET_{locat}$ and $ET_{sota}$ the execution times of a program-input pair tuned by LOCAT and by a SOTA approach, respectively. 

Figure~\ref{fig:speedup} shows the results on the four-node ARM cluster. As can be seen, LOCAT significantly improves the performance of the experimented 25 program-input pairs tuned by other SOTA approaches. In detail, the speedups of the program-input pairs tuned by LOCAT over they tuned by Tuneful, DAC, GBO-RL, and QTune are $2.4\times$, $2.2\times$, $2.0\times$, and $1.9\times$ on average, and up to $3.7\times$, $3.1\times$, $2.8\times$, and $2.4\times$, respectively. Figure~\ref{fig:speedup-8nodes} shows the results on the eight-node x86 cluster where LOCAT still significantly outperforms the SOTA approaches in terms of performance. In detail, LOCAT improves the 25 program-input pairs' performance tuned by Tuneful, DAC, GBO-RL, and QTune by factors of $2.8\times$, $2.6\times$, $2.3\times$, and $2.1\times$ on average, and up to $4.8\times$, $4.7\times$, $3.7\times$, and $3.3\times$, respectively.

A couple of conclusions can be made from these speedups in addition to the optimization time reductions. For one, LOCAT can tune Spark SQL applications with not only higher performance improvements but also in significantly shorter time compared to the SOTA approaches. Second, on significantly different hardware and different scales of clusters, LOCAT can still outperform the SOTA approaches in both performance improvement and optimization time reduction. Third, LOCAT outperforms the SOTA approaches for all different input data sizes of a Spark SQL application, as shown in Figure~\ref{fig:speedup} and Figure~\ref{fig:speedup-8nodes}. Last, LOCAT generally improves the performance more for the larger input data size of a Spark SQL application compared to the SOTA approaches. These benefits make LOCAT more suitable for optimizing future Spark SQL applications because the input data size of them is getting increasingly larger.

\begin{figure}[t!]
  \centering
  \includegraphics[width=2.4in]{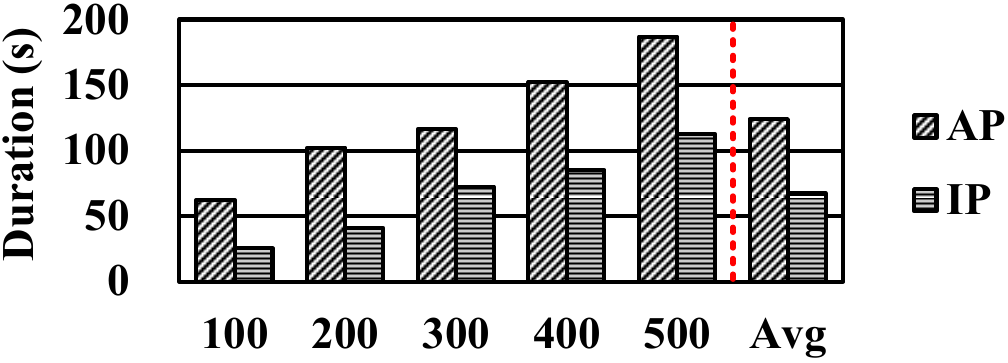}
  \vspace{-4mm}
  \caption{The performance of TPC-DS with input data sizes of 100GB, 200GB, 300GB, 400GB, and 500GB tuned by LOCAT with all parameters (AP) and important parameters  (IP).}\label{fig:LOCATIandQPerf}
  \vspace{-4mm}
\end{figure}

Although LOCAT's primary goal is to reduce the optimization time of ML-based tuning approaches for Spark SQL applications, it surprisingly shows performance improvements. This is because LOCAT identifies the important configuration parameters to tune performance. Tuning more configuration parameters does not necessarily result in higher performance. Instead, it may degrade performance because the unimportant parameters may counteract the performance improvements caused by tuning the important ones. We conduct experiments to confirm this. We compare the performance of TPC-DS with input data sizes of 100GB, 200GB, 300GB, 400GB, and 500GB tuned by LOCAT with all the 38 configuration parameters (AP) and with the 15 important parameters (IP) produced by IICP. Figure~\ref{fig:LOCATIandQPerf} shows the results. As can be seen, The performance achieved by tuning the 15 important parameters is $1.8\times$ higher than that turned by all the 38 parameters on average. This confirms that tuning the important configuration parameters results in higher performance than tuning all the configuration parameters for Spark SQL applications.

\subsection{Why IICP?}
IICP is designed to identify the important configuration parameters with respect to performance. In fact, a lot of machine learning (ML) algorithms can also be used for this purpose. For example, Gradient Boosting Regression Tree (GBRT) has been used to quantify the importance of CPU performance events~\cite{counterminer}. Why do we design IICP for the same purpose? To answer this question, we evaluate if our IICP approach is better than ML-based approaches. 

To identify important configuration parameters, ML-based approaches need to build an accurate performance model as a function of the parameters first. Subsequently, the importance of a parameter is calculated by using the performance model. Higher accuracy of a performance model generally indicates the parameter importance calculated from it is more convincible. We therefore use several ML algorithms to construct performance models and use the mean squared error (MSE)~\cite{marmolin1986subjective} to measure the model accuracy.

\begin{figure}[t!]
  \centering
  \includegraphics[width=\linewidth]{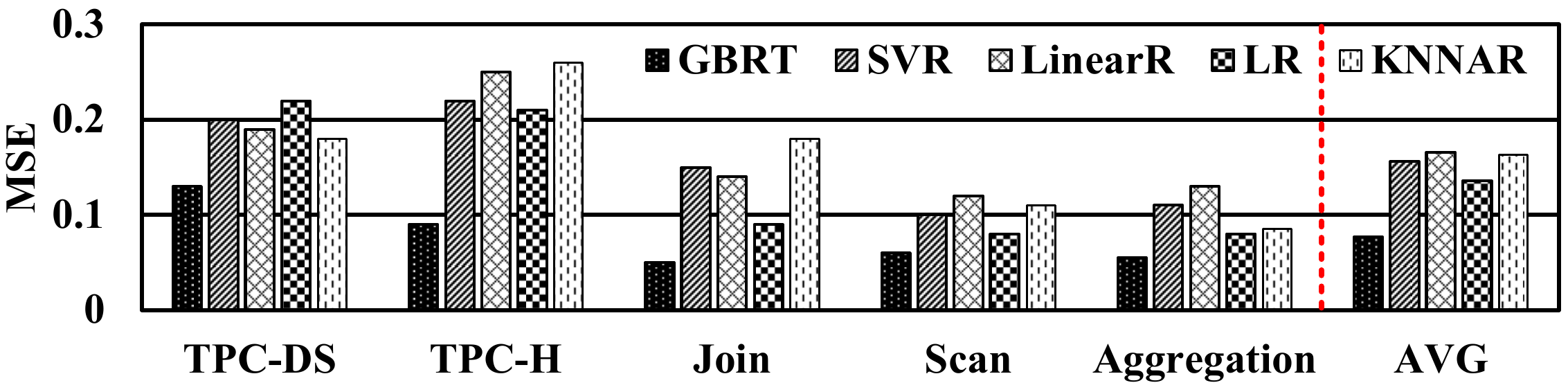}
  \vspace{-6mm}
  \caption{Accuracy of models built by GBRT (Gradient Boosted Regression Tree), SVR (Support Vector Regression), LinearR (Linear Regression), LR (Logistic Regression), and KNNAR (K-Nearest Neighbor Algorithm for Regression).}\label{fig:algorithm comparasion}
  \vspace{-2mm}
\end{figure}

Figure~\ref{fig:algorithm comparasion} shows the accuracy of the performance models built by Gradient Boosting Regression Tree (GBRT), Support Vector Regression (SVR), Linear Regression (LinearR), Logistic Regression (LR), and K-Nearest Neighbor Algorithm for Regression (KNNAR). Note that these models are trained by using the same training data set. As can be seen, the average error of the GBRT models is less than 15\%, which is the lowest among all models built by other ML algorithms for all workloads. This indicates that the configuration parameter importance calculated by GBRT models is more convincible than by other ML algorithms. We therefore compare IICP against the models built by GBRT.

After we identify the important configuration parameters by our IICP and the GBRT model, we use them to configure $TPC-DS$ with 100GB of input data and execute the program a number of times, each time with a different random configuration. Note that the configurations only contain the values of the identified important parameters. We run $TPC-DS$ for $5$, $10$, $15$, $20$, $25$, and $30$ times and observe the standard deviations of its execution times. 

\begin{figure}[t!]
  \centering
  \includegraphics[width=3.3in]{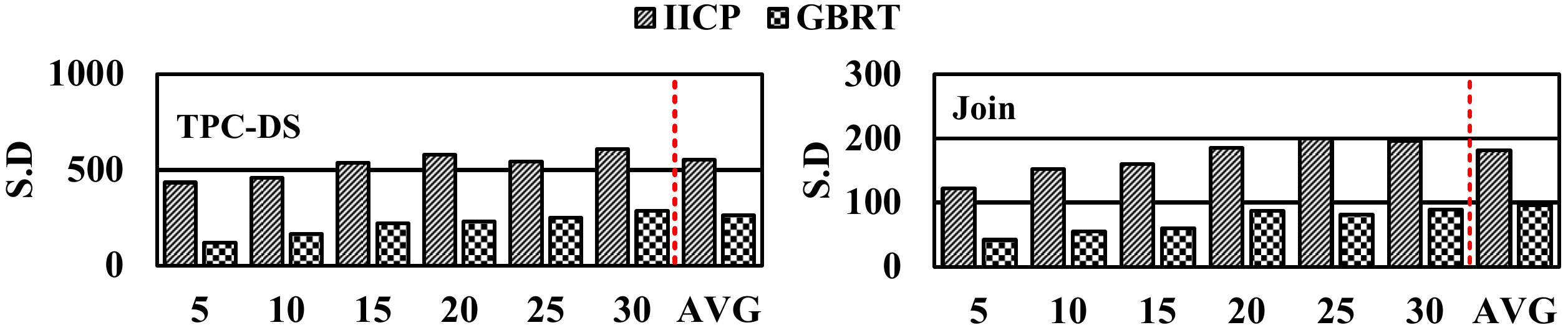}
  \vspace{-2mm}
  \caption{The comparison between IICP and GBRT. The Y axis represents the standard deviation of the execution times of $TPC-DS$ and $Join$ configured by the important parameters identified by IICP or GBRT.}\label{fig:IICPandGBRT}
  \vspace{-4mm}
\end{figure}

The results are shown in Figure~\ref{fig:IICPandGBRT}, where the $Y$ axis represents the SD (standard deviation) of execution times of $TPC-DS$ and {\it Join} configured by the important parameters identified by IICP and GBRT, and the $X$ axis represents the number of executions. Higher SD of execution times indicates that configuration parameters identified by the approach are more important than one another. We find that, the SD of IICP is significantly higher than that of GBRT. This indicates that IICP outperforms GBRT for identifying important parameters, especially with low overhead. The reason is that GBRT requires a large number of experiment samples to build an accurate model while IICP does not, and IICP employs a {\it novel hybrid} approach combining the feature selection and feature extraction.

\subsection{Where does the Speedup Come from?}
\begin{figure}[t!]
  \centering
  \includegraphics[width=\linewidth]{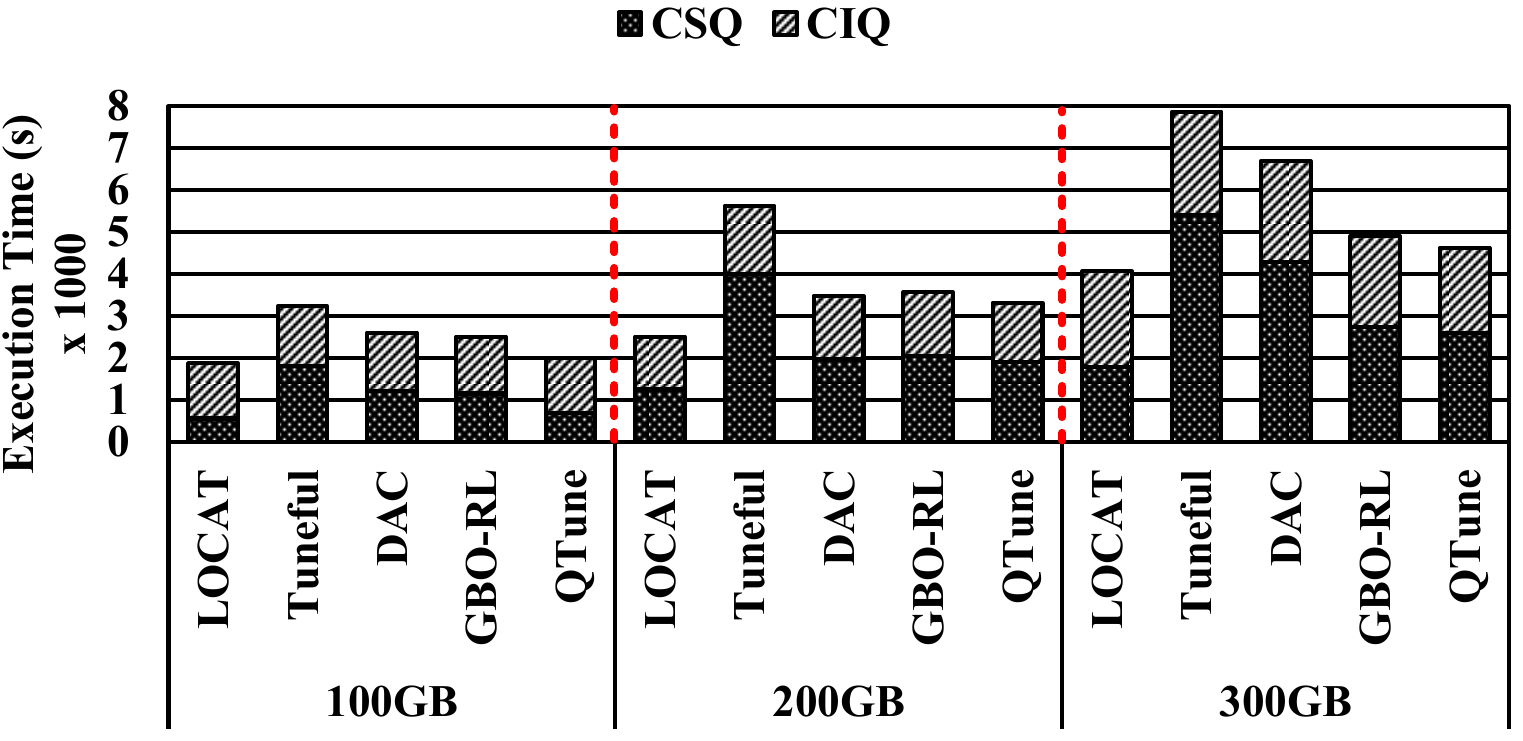}
  \vspace{-6mm}
  \caption{LOCAT Outperforms other Approaches by Significantly Accelerating the Execution of CSQ.}
  \label{fig:IICPQIAcom}
  \vspace{-2mm}
\end{figure}

We now analyze where the speedup made by LOCAT come from. Figure~\ref{fig:IICPQIAcom} shows the execution time of CSQ (Configuration Sensitive Query) and CIQ (Configuration Insensitive Query) of $TPC-DS$ including $104$ queries. A number of interesting observations can be made here. First, LOCAT, Tuneful, DAC, GBO-RL, and QTune all reduce the execution time significantly and the performance is higher with larger input data size. Second, the performance improvement mainly comes from reducing the execution time of CSQ. This is because CIQ is hard to optimize by tuning the configuration parameters. Third, LOCAT outperforms other four methods in reducing more executing time of CSQ. This is because: 1) LOCAT leverages IICP to identify the important parameters and focuses on tuning them, which achieves higher performance improvements as shown in Section~\ref{sec:speedup}; 2) LOCAT distinguishes CIQ between CSQ and avoids executing CIQ to concentrate more on CSQ tuning while other methods is not able to do so; 3) LOCAT takes the size of input data as a prior knowledge of BO to determine current optimal parameter online during the iteration of BO. 

\begin{figure}
  \centering
  \includegraphics[width=\linewidth]{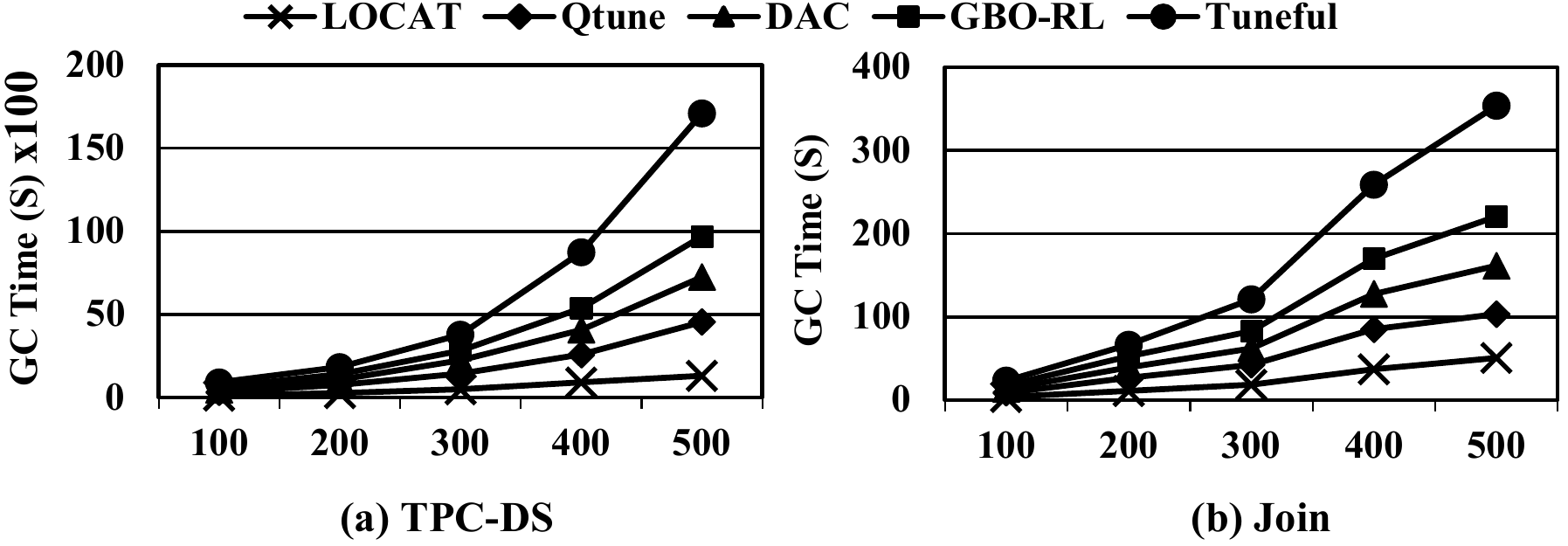}
  \vspace{-6mm}
  \caption{JVM Garbage Collection (GC) Time Comparison.}
  \label{fig:GCTime}
  \vspace{-4mm}
\end{figure}

Furthermore, we find that the speedup made by LOCAT over other SOTA approaches mainly comes from that LOCAT reduces the garbage collection (GC) time significantly more than other approaches. Figure~\ref{fig:GCTime} (a) and (b) show the GC time comparison for $TPC-DS$ with multiple queries and {\it Join} with one query, respectively. As can be seen, the JVM GC time used by LOCAT is significantly shorter than other approaches no matter in with multiple queries or with only one query. In addition, the GC time used by LOCAT increases significantly slowly than other approaches with the increasing of input data size. This is because GC dynamically performs several memory operations such as allocating from or releasing memory to the operating system according to an application's request. LOCAT sets more proper values for the memory related parameters, making GC spend less time to perform the memory operations than other approaches. This also indicates that LOCAT would have more benefits if larger data set is processed.

\subsection{Tuning Overhead of Increasing Data Size}

\begin{figure}[t!]
  \centering
  \includegraphics[width=\linewidth]{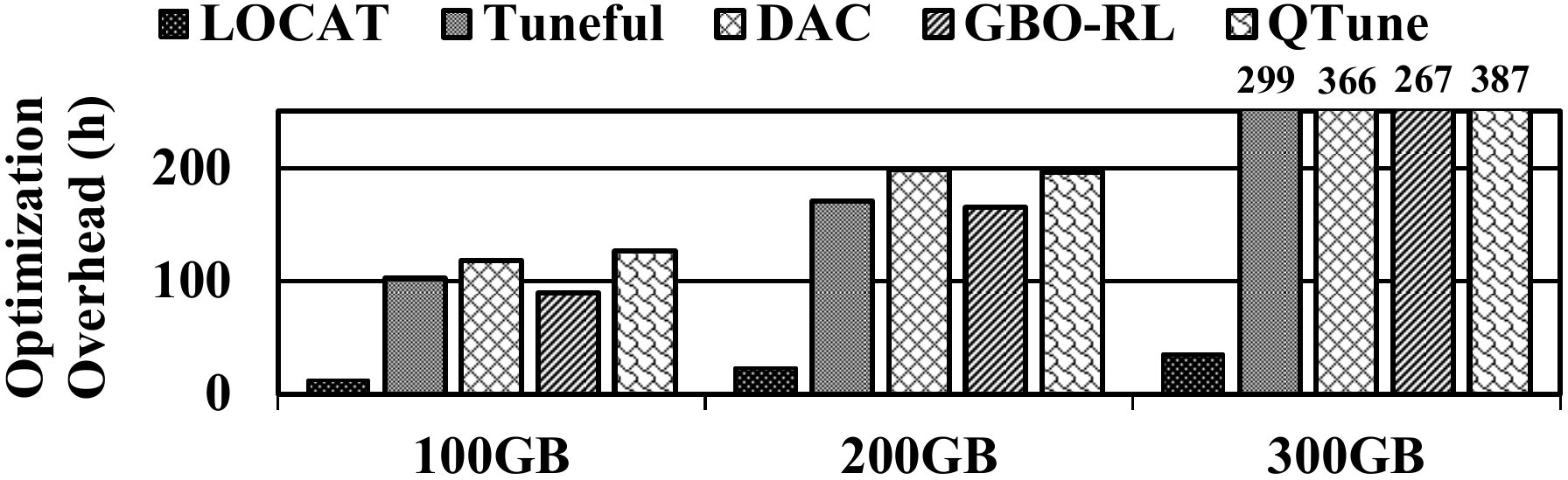}
  \vspace{-6mm}
  \caption{How Tuning Overhead (in hours) Changes when the input data size of an application increases.}
  \label{fig:overheadDataSize}
  \vspace{-4mm}
\end{figure}

Figure~\ref{fig:overheadDataSize} compares the optimization overhead when LOCAT and the SOTA approaches are applied to TPC-DS with increasing input data size. As can be seen, LOCAT incurs significantly lower optimization overhead for all different sizes of input data. The optimization time reduction compared to other SOTA applications made by LOCAT becomes more when the data size increases. The reason is that LOCAT adapts to the input data size changes to avoid re-tuning, which is similar to but better than Tuneful that directly uses BO without adapting data size changes, and GBO-RL that leverages a more time-consuming reinforcement learning (RL).

\begin{figure}[t!]
  \centering
  \includegraphics[width=\linewidth]{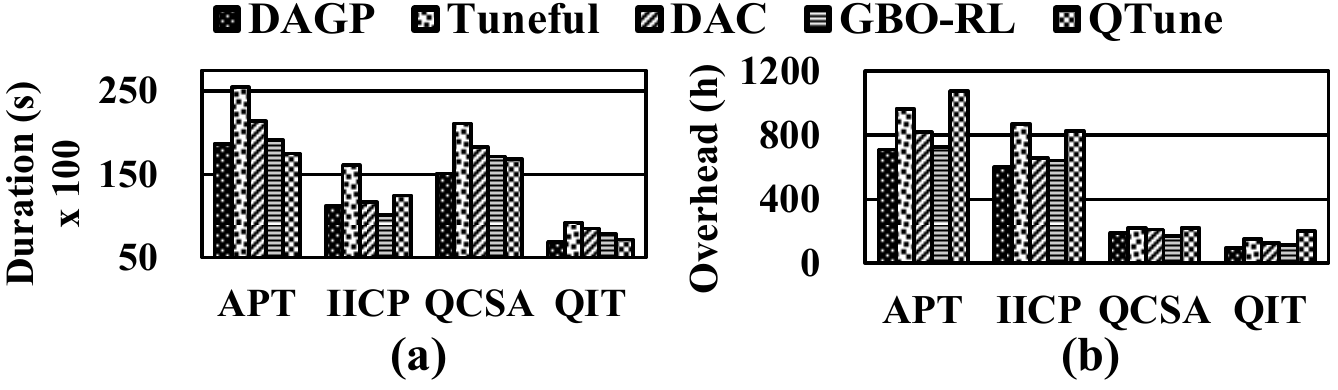}
  \vspace{-8mm}
  \caption{Optimized Performance and Optimization Overhead Comparison between DAGP, Tuneful, DAC, GBO-RL, and Qtune. APT represents all parameters tuning. QIT denotes tuning performance with QCSA and IICP.}
  \label{fig:QI}
  \vspace{-6mm}
\end{figure}

\subsection{Using IICP/QCSA on SOTA Approaches}
Although QCSA and IICP are designed to work with DAGP, they can definitely be applied on other SOTA approaches. We therefore combine Tuneful, DAC, GBO-RL, and QTune with all parameter tuning (APT), QCSA, and IICP to form new approaches first. We subsequently apply these approaches on $TPC-DS$ with 500GB of input data to compare the optimized performance and the optimization overhead. Figure~\ref{fig:QI} shows the result. As can be seen, IICP reduces the execution times of the five approaches by factors of $1.7\times$ on average and up to $1.9\times$ (Figure~\ref{fig:QI} (a)), and decreases the optimization overhead by factors of $1.2\times$ on average and up to $1.3\times$ (Figure~\ref{fig:QI} (b)). QCSA improves the performance of TPC-DS tuned by the five approaches by factors of $1.3\times$ on average and up to $1.4\times$ and reduces their optimization overhead by factors of $4.2\times$ and up to $4.8\times$. Moreover, we find that IICP with QCSA can further improve the performance by a factor of $2.6\times$ on average and up to $2.9\times$, and decreases the optimization overhead by a factor of $6.8\times$ on average and up to $9.1\times$. These results indicate that our IICP and QCSA techniques can be used in other ML approaches to tune the performance with low overhead. 

\subsection{Reasons for Config In/Sensitive Queries}\label{sec:rcsqciq}
We now analyze why some queries are configuration sensitive while others are not. According to \cite{pavlo2009comparison}, queries can be divided into three categories: ‘selection’, ‘join’, and ‘aggregation’. Most simple ‘selection’ queries are configuration insensitive because they do not consume a large amount of computing and memory resources specified by the configuration parameters. For example, in TPC-DS with 100GB of input data, {\tt \{Q09, Q13, Q16, Q28, Q32, Q38, Q48, Q61, Q84, Q87, Q88, Q94, and Q96\}} are 'selection' queries which perform a simple filter logic and only consume 5 CPU cores and 8GB memory on average to achieve their optimal performance. Tuning the resource related configuration parameters for them therefore does not influence their performance significantly. Moreover, these queries do not process or move a large amount of data. The configuration parameter related to “compress” data therefore does not affect the performance significantly.

However, the queries belonging to categories ‘join’ and ‘aggregation’ involve shuffle operations which generally consume compute, memory, and network resources specified by configuration parameters. If the shuffle operations of a query process a large amount of data, the query is configuration sensitive. Otherwise, the query is configuration insensitive. For example, in $TPC-DS$ with 100GB of input data, the shuffle operations of $Q72$ process 52GB of data. $Q72$ is therefore configuration sensitive. In contrast, the shuffle operations of $Q08$ process only 5MB of data, $Q08$ is therefore insensitive.

\subsection{Parameter Value Range Determination}\label{sec:vrd}
When we tune configuration parameters for optimal performance of a Spark SQL application, it is important to know the value range of each parameter. However, this is difficult because the value ranges of some parameters depend on the cluster resources while others depend on the program logic. To address these difficulties, we firstly classify the 38 configuration parameters into 28 numeric parameters (in bold text in Table~\ref{table:parameters}) and 10 non-numeric ones (others in Table~\ref{table:parameters}). Since the values of non-numeric parameters are always {\it True} or {\it False}, we do not determine the value ranges for them and instead focus on the numeric parameters. 

Subsequently, we further classify the numerical parameters into resource (e.g., CPU core \# and memory size) parameters (with $*$ in Table~\ref{table:parameters}), and non-resource parameters (in bold text without $*$ in Table~\ref{table:parameters}). Next, we determine the value range of a resource parameter according to  the memory size and core \# in total in the cluster and the maximum resource capacity of the container specified by cluster managers such as Yarn~\cite{vavilapalli2013yarn}, Kubernetes~\cite{bernstein2014containers}, and Mesos~\cite{hindman2011mesos}. For example, in Yarn mode, the value range of {\it spark.executor.cores} and {\it spark.driver.cores} is the same as that of CPU core capacity of Yarn container. As for memory related parameters, the value range of {\it spark.driver.memory} and {\it spark.executor.memory} is the same as that of memory capacity of Yarn container. The value range of {\it spark.executor.memoryOverhead} and {\it spark.memory.offHeap.size} is from 0 to the maximum of memory capacity of Yarn container, which is determined by time-consuming experiments. Moreover, we control the sum of parameters {\it spark.executor.memory}, {\it spark.executor.\\memoryOverhead}, and {\it spark.memory.offHeap.size}, the total memory resource of a single process, to be smaller than the memory capacity of the Yarn container. In addition, we also specify the product of {\it spark.executor.instances} and {\it the resource amount of a single process} to be less than the total amount of resources in the cluster. As such, the value ranges of the resource parameters can be determined.

As for the non-resource parameters, we determine their value range based on a time-consuming experiment. Starting from the default value ($dv$) of a parameter, we try values larger or smaller than the $dv$ with $N$ times. The value of $N^{th}$ time is $dv+N*step$ when we try to determine the upper bound. Likewise, the value is $dv-N*step$ when we try to determine the lower bound. $step$ is the stride value. If a program can not start to run or hung when we configure the value of a parameter as $dv+N*step$ or $dv-N*step$, we set the upper bound or lower bound of the parameter as $dv+(N-1)*step$ or $dv-(N-1)*step$, respectively. 

\section{Related Work}\label{sec:rlw}
In this section, we describe the configuration auto-tuning studies related to Spark SQL. A large body of automatic parameter tuning approaches can be applied to Spark SQL, which are divided into six categories~\cite{herodotou2020survey}: 1) \textbf{Rule-based approaches (RBA)} tune performance based on the expert experience, online tutorials~\cite{SparkConfWebsite}, or tuning guidebooks~\cite{SparkConfTuningWebsite} which are time-consuming and labour-intensive because using RBA requires a deep understanding of system internals, while LOCAT does not. 2) \textbf{Cost modeling approaches' (CMA)}~\cite{wang2015performance, venkataraman2016ernest, zacheilas2017dione, singhal2017performance, gounaris2017dynamic, chen2019cost} build performance prediction models with analytical model which are not able to be applied to the complex system like Spark SQL and adapt to the input data size changing, while LOCAT is able to. 3) \textbf{Simulation-based approaches (SBA)}~\cite{ardagna2020predicting, de2019multi, de2018bigdatanetsim, karimian2019scalable} build performance prediction models based on simulation of optimized system which are not suitable for complex system like Spark SQL. 4) \textbf{Experiment-driven approaches (EDA)}~\cite{Petridis2017Trial, zhu2017bestconfig, gounaris2017dynamic, bao2018learning, yu2018datasize} find the optimal configuration by executing an application repeatedly with different configuration parameters until converge, which causes high overhead, while LOCAT really takes optimization overhead into consideration and significantly reduces it. 5) \textbf{Machine learning approaches (MLA)}~\cite{wang2016novel, hernandez2018using, jia2016auto, chen2019d} build performance prediction models by machine learning algorithms, needing to collect a large number of training samples with high overhead, while LOCAT achieves high optimization performance with significantly low overhead. 6) \textbf{Adaptive approaches (AA)}~\cite{fekry2020tuneful, suraweera2002kermit, li2019qtune, kunjir2020black} tune the configuration parameter with adaptivity to dynamic runtime status (e.g., input data size changing). AA does not consider the optimization overhead while LOCAT does.

From above, we can find that current approaches still face two problems: First, high overhead of the optimization process. Second, unadaptability of optimal configuration to different input data sizes. We now summarize studies trying to solve the two problems.

\subsection{Reducing Optimization Overhead} 
Leveraging ML-based performance models to avoid actually executing an application is an effective way to reduce the cost of searching for the optimal configuration for the application. RFHOC~\cite{rfhoc} and DAC~\cite{yu2018datasize} are the examples of such approaches. However, it is time-consuming to build a performance model because we need to collect a large number of training samples by executing the application on a real cluster with a random configuration each time. LOCAT differs from these approaches by innovating QCSA and IICP to reduce the time used for collecting training samples.

Another line of studies to accelerate the optimization process are to combine an analytical model or the application characteristics with a ML algorithm. For example, GBO-RL~\cite{kunjir2020black} builds an analytical model for Spark's memory management to speed up the BO convergence. But GBO-RL only considers memory and the analytical model is inaccurate~\cite{yu2018datasize}. In contrast, LOCAT considers all layers of configuration parameters of Spark SQL applications. Another work, Tuneful~\cite{fekry2020tuneful}, uses One-at-a-time (OAT) method to identify a subspace composed of configurations that significantly influence Spark performance as the searching space, which directly reduces the dimension of the search space to speed the optimization process up. But Tuneful is not suitable for high-dimensional configuration scenarios because the number of iterations of OAT increases rapidly when the number of configuration parameters increases. 

\subsection{Adapting to Input Data Size Changes} 
Some performance tuning studies for big data systems try to adapt to input data size changes indirectly or directly. For instance, KERMIT~\cite{suraweera2002kermit} works at YARN level to dynamically adapt to the resource allocation, including memory and CPU cores of containers. It randomly searches an optimal resource configuration by observing the container's performance changing in real-time. But KERMIT is limited to optimize memory configuration only and the random searching is always time-consuming. \cite{prats2020you} is able to optimize the unseen workload without retraining by leveraging a performance model enhanced by using the execution log. It generalizes to different workloads (with different logical or input data) by extracting information in stages and tasks to provide a general model for all kinds of applications in Spark. But \cite{prats2020you} becomes impractical when there is a shortage of log files. QTune~\cite{li2019qtune} realizes adaptive optimization by leveraging deep reinforcement learning~\cite{sutton2018reinforcement}, and QTune needs to characterize each query of an application for high performance. CDBTune~\cite{zhang2019end} is similar to QTune, leveraging deep reinforcement learning. They both are too time-consuming to be applied in practice.

Differing from these studies, LOCAT realizes adaptive optimization for Spark SQL applications by developing a Data size Aware Gaussian Process (DAGP) which considers the input data size in Gaussian Process. Moreover, LOCAT significantly reduces the optimization overhead and outperforms these approaches in high-dimensional configuration spaces.

\section{Conclusion}\label{sec:con}
This paper proposes LOCAT, a BO-based approach that efficiently as well as adaptively finds the optimal configurations to achieve high performance for a Spark SQL application on a given cluster. LOCAT innovates three techniques: query configuration sensitive analysis (QCSA), identifying important configuration parameters (IICP), and data size aware Gaussian Process (DAGP). The experiments on two significantly different clusters, a four-node ARM cluster and an eight-node x86 cluster, show that LOCAT can significantly reduce the optimization time of the state-of-the-art approaches and dramatically improve performance over them.

\section{Acknowledgements}
This work is supported by the Key-Area Research and Development Program of Guangdong Province (Grant No. 2019B010155003), and is partially supported by the Shenzhen Institute of Artificial Intelligence and Robotics for Society (AIRS), Chinese University of Hong Kong, Shenzhen. 

\bibliographystyle{ACM-Reference-Format}
\bibliography{references}


\begin{thebibliography}{74}


\ifx \showCODEN    \undefined \def \showCODEN     #1{\unskip}     \fi
\ifx \showDOI      \undefined \def \showDOI       #1{#1}\fi
\ifx \showISBNx    \undefined \def \showISBNx     #1{\unskip}     \fi
\ifx \showISBNxiii \undefined \def \showISBNxiii  #1{\unskip}     \fi
\ifx \showISSN     \undefined \def \showISSN      #1{\unskip}     \fi
\ifx \showLCCN     \undefined \def \showLCCN      #1{\unskip}     \fi
\ifx \shownote     \undefined \def \shownote      #1{#1}          \fi
\ifx \showarticletitle \undefined \def \showarticletitle #1{#1}   \fi
\ifx \showURL      \undefined \def \showURL       {\relax}        \fi
\providecommand\bibfield[2]{#2}
\providecommand\bibinfo[2]{#2}
\providecommand\natexlab[1]{#1}
\providecommand\showeprint[2][]{arXiv:#2}

\bibitem[Akoglu(2018)]%
        {akoglu2018user}
\bibfield{author}{\bibinfo{person}{Haldun Akoglu}.}
  \bibinfo{year}{2018}\natexlab{}.
\newblock \showarticletitle{User's guide to correlation coefficients}.
\newblock \bibinfo{journal}{\emph{Turkish Journal of Emergency Medicine}}
  \bibinfo{volume}{18}, \bibinfo{number}{3} (\bibinfo{year}{2018}),
  \bibinfo{pages}{91--93}.
\newblock
\showISSN{2452-2473}
\urldef\tempurl%
\url{https://doi.org/10.1016/j.tjem.2018.08.001}
\showDOI{\tempurl}


\bibitem[Alipourfard et~al\mbox{.}(2017)]%
        {alipourfard2017cherrypick}
\bibfield{author}{\bibinfo{person}{Omid Alipourfard},
  \bibinfo{person}{Hongqiang~Harry Liu}, \bibinfo{person}{Jianshu Chen},
  \bibinfo{person}{Shivaram Venkataraman}, \bibinfo{person}{Minlan Yu}, {and}
  \bibinfo{person}{Ming Zhang}.} \bibinfo{year}{2017}\natexlab{}.
\newblock \showarticletitle{Cherrypick: Adaptively Unearthing the Best Cloud
  Configurations for Big Data Analytics}. In
  \bibinfo{booktitle}{\emph{Proceedings of the 14th USENIX Conference on
  Networked Systems Design and Implementation}} (Boston, MA, USA)
  \emph{(\bibinfo{series}{NSDI'17})}. \bibinfo{publisher}{USENIX Association},
  \bibinfo{address}{USA}, \bibinfo{pages}{469–482}.
\newblock
\showISBNx{9781931971379}
\urldef\tempurl%
\url{https://www.usenix.org/conference/nsdi17/technical-sessions/presentation/alipourfard}
\showURL{%
\tempurl}


\bibitem[Ardagna et~al\mbox{.}(2021)]%
        {ardagna2020predicting}
\bibfield{author}{\bibinfo{person}{D. Ardagna}, \bibinfo{person}{E.
  Barbierato}, \bibinfo{person}{E. Gianniti}, \bibinfo{person}{M. Gribaudo},
  \bibinfo{person}{T.~B.M. Pinto}, \bibinfo{person}{A.~P.C. da Silva}, {and}
  \bibinfo{person}{J.~M. Almeida}.} \bibinfo{year}{2021}\natexlab{}.
\newblock \showarticletitle{{Predicting the performance of big data
  applications on the cloud}}.
\newblock \bibinfo{journal}{\emph{Journal of Supercomputing}}
  \bibinfo{volume}{77}, \bibinfo{number}{2} (\bibinfo{year}{2021}),
  \bibinfo{pages}{1321--1353}.
\newblock
\showISSN{15730484}
\urldef\tempurl%
\url{https://doi.org/10.1007/s11227-020-03307-w}
\showDOI{\tempurl}


\bibitem[Armbrust et~al\mbox{.}(2015)]%
        {armbrust2015spark}
\bibfield{author}{\bibinfo{person}{Michael Armbrust},
  \bibinfo{person}{Reynold~S. Xin}, \bibinfo{person}{Cheng Lian},
  \bibinfo{person}{Yin Huai}, \bibinfo{person}{Davies Liu},
  \bibinfo{person}{Joseph~K. Bradley}, \bibinfo{person}{Xiangrui Meng},
  \bibinfo{person}{Tomer Kaftan}, \bibinfo{person}{Michael~J. Franklin},
  \bibinfo{person}{Ali Ghodsi}, {and} \bibinfo{person}{Matei Zaharia}.}
  \bibinfo{year}{2015}\natexlab{}.
\newblock \showarticletitle{Spark SQL: Relational Data Processing in Spark}. In
  \bibinfo{booktitle}{\emph{Proceedings of the 2015 ACM SIGMOD International
  Conference on Management of Data}} (Melbourne, Victoria, Australia)
  \emph{(\bibinfo{series}{SIGMOD '15})}. \bibinfo{publisher}{Association for
  Computing Machinery}, \bibinfo{address}{New York, NY, USA},
  \bibinfo{pages}{1383–1394}.
\newblock
\showISBNx{9781450327589}
\urldef\tempurl%
\url{https://doi.org/10.1145/2723372.2742797}
\showDOI{\tempurl}


\bibitem[Baldacci and Golfarelli(2019)]%
        {baldacci2018cost}
\bibfield{author}{\bibinfo{person}{Lorenzo Baldacci} {and}
  \bibinfo{person}{Matteo Golfarelli}.} \bibinfo{year}{2019}\natexlab{}.
\newblock \showarticletitle{A Cost Model for SPARK SQL}.
\newblock \bibinfo{journal}{\emph{IEEE Transactions on Knowledge and Data
  Engineering}} \bibinfo{volume}{31}, \bibinfo{number}{5}
  (\bibinfo{year}{2019}), \bibinfo{pages}{819--832}.
\newblock
\urldef\tempurl%
\url{https://doi.org/10.1109/TKDE.2018.2850339}
\showDOI{\tempurl}


\bibitem[Bao et~al\mbox{.}(2018)]%
        {bao2018learning}
\bibfield{author}{\bibinfo{person}{Liang Bao}, \bibinfo{person}{Xin Liu}, {and}
  \bibinfo{person}{Weizhao Chen}.} \bibinfo{year}{2018}\natexlab{}.
\newblock \showarticletitle{Learning-based Automatic Parameter Tuning for Big
  Data Analytics Frameworks}. In \bibinfo{booktitle}{\emph{2018 IEEE
  International Conference on Big Data (Big Data)}}. \bibinfo{pages}{181--190}.
\newblock
\urldef\tempurl%
\url{https://doi.org/10.1109/BigData.2018.8622018}
\showDOI{\tempurl}


\bibitem[Barata et~al\mbox{.}(2015)]%
        {barata2015overview}
\bibfield{author}{\bibinfo{person}{Melyssa Barata}, \bibinfo{person}{Jorge
  Bernardino}, {and} \bibinfo{person}{Pedro Furtado}.}
  \bibinfo{year}{2015}\natexlab{}.
\newblock \showarticletitle{{An overview of decision support benchmarks:
  TPC-DS, TPC-H and SSB}}.
\newblock \bibinfo{journal}{\emph{Advances in Intelligent Systems and
  Computing}}  \bibinfo{volume}{353} (\bibinfo{year}{2015}),
  \bibinfo{pages}{619--628}.
\newblock
\showISBNx{9783319164854}
\showISSN{21945357}
\urldef\tempurl%
\url{https://doi.org/10.1007/978-3-319-16486-1_61}
\showDOI{\tempurl}


\bibitem[Bei et~al\mbox{.}(2016)]%
        {rfhoc}
\bibfield{author}{\bibinfo{person}{Zhendong Bei}, \bibinfo{person}{Zhibin Yu},
  \bibinfo{person}{Huiling Zhang}, \bibinfo{person}{Wen Xiong},
  \bibinfo{person}{Chengzhong Xu}, \bibinfo{person}{Lieven Eeckhout}, {and}
  \bibinfo{person}{Shengzhong Feng}.} \bibinfo{year}{2016}\natexlab{}.
\newblock \showarticletitle{RFHOC: A Random-Forest Approach to Auto-Tuning
  Hadoop's Configuration}.
\newblock \bibinfo{journal}{\emph{IEEE Transactions on Parallel and Distributed
  Systems}} \bibinfo{volume}{27}, \bibinfo{number}{5} (\bibinfo{year}{2016}),
  \bibinfo{pages}{1470--1483}.
\newblock
\urldef\tempurl%
\url{https://doi.org/10.1109/TPDS.2015.2449299}
\showDOI{\tempurl}


\bibitem[Bernstein(2014)]%
        {bernstein2014containers}
\bibfield{author}{\bibinfo{person}{David Bernstein}.}
  \bibinfo{year}{2014}\natexlab{}.
\newblock \showarticletitle{Containers and Cloud: From LXC to Docker to
  Kubernetes}.
\newblock \bibinfo{journal}{\emph{IEEE Cloud Computing}} \bibinfo{volume}{1},
  \bibinfo{number}{3} (\bibinfo{year}{2014}), \bibinfo{pages}{81--84}.
\newblock
\urldef\tempurl%
\url{https://doi.org/10.1109/MCC.2014.51}
\showDOI{\tempurl}


\bibitem[Bhattacharyya et~al\mbox{.}(1979)]%
        {kleinbaum2013applied}
\bibfield{author}{\bibinfo{person}{Helen~T. Bhattacharyya},
  \bibinfo{person}{David~G. Kleinbaum}, {and} \bibinfo{person}{Lawrence~L.
  Kupper}.} \bibinfo{year}{1979}\natexlab{}.
\newblock \bibinfo{booktitle}{\emph{{Applied Regression Analysis and Other
  Multivariable Methods.}}} Vol.~\bibinfo{volume}{74}.
\newblock \bibinfo{publisher}{Cengage Learning}. 732 pages.
\newblock
\showISSN{01621459}
\urldef\tempurl%
\url{https://doi.org/10.2307/2287012}
\showDOI{\tempurl}


\bibitem[Bolboaca and J{\"a}ntschi(2006)]%
        {bolboaca2006pearson}
\bibfield{author}{\bibinfo{person}{Sorana-Daniela Bolboaca} {and}
  \bibinfo{person}{Lorentz J{\"a}ntschi}.} \bibinfo{year}{2006}\natexlab{}.
\newblock \showarticletitle{Pearson versus Spearman, Kendall’s tau
  correlation analysis on structure-activity relationships of biologic active
  compounds}.
\newblock \bibinfo{journal}{\emph{Leonardo Journal of Sciences}}
  \bibinfo{volume}{5}, \bibinfo{number}{9} (\bibinfo{year}{2006}),
  \bibinfo{pages}{179--200}.
\newblock
\urldef\tempurl%
\url{http://ljs.academicdirect.org/A09/179_200.htm}
\showURL{%
\tempurl}


\bibitem[Boncz et~al\mbox{.}(2014)]%
        {boncz2013tpc}
\bibfield{author}{\bibinfo{person}{Peter Boncz}, \bibinfo{person}{Thomas
  Neumann}, {and} \bibinfo{person}{Orri Erling}.}
  \bibinfo{year}{2014}\natexlab{}.
\newblock \showarticletitle{{TPC-H analyzed: Hidden messages and lessons
  learned from an influential benchmark}}. In \bibinfo{booktitle}{\emph{Lecture
  Notes in Computer Science (including subseries Lecture Notes in Artificial
  Intelligence and Lecture Notes in Bioinformatics)}},
  Vol.~\bibinfo{volume}{8391 LNCS}. Springer, \bibinfo{pages}{61--76}.
\newblock
\showISBNx{9783319049359}
\showISSN{16113349}
\urldef\tempurl%
\url{https://doi.org/10.1007/978-3-319-04936-6_5}
\showDOI{\tempurl}


\bibitem[Chandrashekar and Sahin(2014)]%
        {chandrashekar2014survey}
\bibfield{author}{\bibinfo{person}{Girish Chandrashekar} {and}
  \bibinfo{person}{Ferat Sahin}.} \bibinfo{year}{2014}\natexlab{}.
\newblock \showarticletitle{A survey on feature selection methods}.
\newblock \bibinfo{journal}{\emph{Computers \& Electrical Engineering}}
  \bibinfo{volume}{40}, \bibinfo{number}{1} (\bibinfo{year}{2014}),
  \bibinfo{pages}{16--28}.
\newblock
\showISSN{0045-7906}
\urldef\tempurl%
\url{https://doi.org/10.1016/j.compeleceng.2013.11.024}
\showDOI{\tempurl}
\newblock
\shownote{40th-year commemorative issue}.


\bibitem[Chen et~al\mbox{.}(2019a)]%
        {chen2019d}
\bibfield{author}{\bibinfo{person}{Yuxing Chen}, \bibinfo{person}{Peter
  Goetsch}, \bibinfo{person}{Mohammad~A. Hoque}, \bibinfo{person}{Jiaheng Lu},
  {and} \bibinfo{person}{Sasu Tarkoma}.} \bibinfo{year}{2019}\natexlab{a}.
\newblock \showarticletitle{{d-Simplexed: Adaptive Delaunay Triangulation for
  Performance Modeling and Prediction on Big Data Analytics}}.
\newblock \bibinfo{journal}{\emph{IEEE Transactions on Big Data}}
  (\bibinfo{year}{2019}), \bibinfo{pages}{1--1}.
\newblock
\urldef\tempurl%
\url{https://doi.org/10.1109/tbdata.2019.2948338}
\showDOI{\tempurl}


\bibitem[Chen et~al\mbox{.}(2019b)]%
        {chen2019cost}
\bibfield{author}{\bibinfo{person}{Yuxing Chen}, \bibinfo{person}{Jiaheng Lu},
  \bibinfo{person}{Chen Chen}, \bibinfo{person}{Mohammad Hoque}, {and}
  \bibinfo{person}{Sasu Tarkoma}.} \bibinfo{year}{2019}\natexlab{b}.
\newblock \showarticletitle{{Cost-effective resource provisioning for spark
  workloads}}. In \bibinfo{booktitle}{\emph{International Conference on
  Information and Knowledge Management, Proceedings}}.
  \bibinfo{pages}{2477--2480}.
\newblock
\showISBNx{9781450369763}
\urldef\tempurl%
\url{https://doi.org/10.1145/3357384.3358090}
\showDOI{\tempurl}


\bibitem[Cheng et~al\mbox{.}(2021)]%
        {cheng2021nonlinear}
\bibfield{author}{\bibinfo{person}{Keyang Cheng},
  \bibinfo{person}{Muhammad~Saddam Khokhar}, \bibinfo{person}{Misbah Ayoub},
  {and} \bibinfo{person}{Zakria Jamali}.} \bibinfo{year}{2021}\natexlab{}.
\newblock \showarticletitle{Nonlinear dimensionality reduction in robot vision
  for industrial monitoring process via deep three dimensional Spearman
  correlation analysis (D3D-SCA)}.
\newblock \bibinfo{journal}{\emph{Multimedia Tools and Applications}}
  \bibinfo{volume}{80}, \bibinfo{number}{4} (\bibinfo{year}{2021}),
  \bibinfo{pages}{5997--6017}.
\newblock
\urldef\tempurl%
\url{https://doi.org/10.1007/s11042-020-09859-6}
\showDOI{\tempurl}


\bibitem[Chiba and Onodera(2016)]%
        {chiba2016workload}
\bibfield{author}{\bibinfo{person}{Tatsuhiro Chiba} {and}
  \bibinfo{person}{Tamiya Onodera}.} \bibinfo{year}{2016}\natexlab{}.
\newblock \showarticletitle{{Workload characterization and optimization of
  TPC-H queries on Apache Spark}}. In \bibinfo{booktitle}{\emph{ISPASS 2016 -
  International Symposium on Performance Analysis of Systems and Software}}.
  IEEE, \bibinfo{pages}{112--121}.
\newblock
\showISBNx{9781509019526}
\urldef\tempurl%
\url{https://doi.org/10.1109/ISPASS.2016.7482079}
\showDOI{\tempurl}


\bibitem[Chiba et~al\mbox{.}(2018)]%
        {chiba2018towards}
\bibfield{author}{\bibinfo{person}{Tatsuhiro Chiba}, \bibinfo{person}{Takeshi
  Yoshimura}, \bibinfo{person}{Michihiro Horie}, {and} \bibinfo{person}{Hiroshi
  Horii}.} \bibinfo{year}{2018}\natexlab{}.
\newblock \showarticletitle{{Towards Selecting Best Combination of
  SQL-on-Hadoop Systems and JVMs}}. In \bibinfo{booktitle}{\emph{IEEE
  International Conference on Cloud Computing, CLOUD}},
  Vol.~\bibinfo{volume}{2018-July}. IEEE, \bibinfo{pages}{245--252}.
\newblock
\showISBNx{9781538672358}
\showISSN{21596190}
\urldef\tempurl%
\url{https://doi.org/10.1109/CLOUD.2018.00038}
\showDOI{\tempurl}


\bibitem[Cuzzocrea(2015)]%
        {cuzzocrea2015data}
\bibfield{author}{\bibinfo{person}{Alfredo Cuzzocrea}.}
  \bibinfo{year}{2015}\natexlab{}.
\newblock \showarticletitle{Data warehousing and OLAP over Big Data: a survey
  of the state-of-the-art, open problems and future challenges}.
\newblock \bibinfo{journal}{\emph{International Journal of Business Process
  Integration and Management}} \bibinfo{volume}{7}, \bibinfo{number}{4}
  (\bibinfo{year}{2015}), \bibinfo{pages}{372--377}.
\newblock
\urldef\tempurl%
\url{https://doi.org/10.1504/IJBPIM.2015.073665}
\showDOI{\tempurl}
\showeprint{https://www.inderscienceonline.com/doi/pdf/10.1504/IJBPIM.2015.073665}


\bibitem[de~Almeida et~al\mbox{.}(2018)]%
        {de2018bigdatanetsim}
\bibfield{author}{\bibinfo{person}{Leandro~Batista de Almeida},
  \bibinfo{person}{Eduardo~Cunha de Almeida}, \bibinfo{person}{John Murphy},
  \bibinfo{person}{Robson~E. De~Grande}, {and} \bibinfo{person}{Anthony
  Ventresque}.} \bibinfo{year}{2018}\natexlab{}.
\newblock \showarticletitle{BigDataNetSim: A Simulator for Data and Process
  Placement in Large Big Data Platforms}. In \bibinfo{booktitle}{\emph{2018
  IEEE/ACM 22nd International Symposium on Distributed Simulation and Real Time
  Applications (DS-RT)}}. \bibinfo{pages}{1--10}.
\newblock
\urldef\tempurl%
\url{https://doi.org/10.1109/DISTRA.2018.8601018}
\showDOI{\tempurl}


\bibitem[{De Almeida} et~al\mbox{.}(2019)]%
        {de2019multi}
\bibfield{author}{\bibinfo{person}{Leandro~Batista {De Almeida}},
  \bibinfo{person}{Damien Magoni}, \bibinfo{person}{Philip Perry},
  \bibinfo{person}{Eduardo~Cunha {De Almeida}}, \bibinfo{person}{John Murphy},
  {and} \bibinfo{person}{Anthony Ventresque}.} \bibinfo{year}{2019}\natexlab{}.
\newblock \showarticletitle{{Multi-Layer-Mesh: A Novel Topology and SDN-Based
  Path Switching for Big Data Cluster Networks}}. In
  \bibinfo{booktitle}{\emph{IEEE International Conference on Communications}},
  Vol.~\bibinfo{volume}{2019-May}. IEEE, \bibinfo{pages}{1--7}.
\newblock
\showISBNx{9781538680889}
\showISSN{15503607}
\urldef\tempurl%
\url{https://doi.org/10.1109/ICC.2019.8761785}
\showDOI{\tempurl}


\bibitem[Fekry et~al\mbox{.}(2020)]%
        {fekry2020tuneful}
\bibfield{author}{\bibinfo{person}{Ayat Fekry}, \bibinfo{person}{Lucian
  Carata}, \bibinfo{person}{Thomas Pasquier}, \bibinfo{person}{Andrew Rice},
  {and} \bibinfo{person}{Andy Hopper}.} \bibinfo{year}{2020}\natexlab{}.
\newblock \showarticletitle{{Tuneful: An Online Significance-Aware
  Configuration Tuner for Big Data Analytics}}.
\newblock \bibinfo{journal}{\emph{arXiv preprint arXiv:2001.08002}}
  (\bibinfo{year}{2020}).
\newblock
\showeprint[arxiv]{2001.08002}
\urldef\tempurl%
\url{http://arxiv.org/abs/2001.08002}
\showURL{%
\tempurl}


\bibitem[Gounaris et~al\mbox{.}(2017)]%
        {gounaris2017dynamic}
\bibfield{author}{\bibinfo{person}{Anastasios Gounaris},
  \bibinfo{person}{Georgia Kougka}, \bibinfo{person}{Ruben Tous},
  \bibinfo{person}{Carlos~Tripiana Montes}, {and} \bibinfo{person}{Jordi
  Torres}.} \bibinfo{year}{2017}\natexlab{}.
\newblock \showarticletitle{{Dynamic configuration of partitioning in spark
  applications}}.
\newblock \bibinfo{journal}{\emph{IEEE Transactions on Parallel and Distributed
  Systems}} \bibinfo{volume}{28}, \bibinfo{number}{7} (\bibinfo{year}{2017}),
  \bibinfo{pages}{1891--1904}.
\newblock
\showISSN{10459219}
\urldef\tempurl%
\url{https://doi.org/10.1109/TPDS.2017.2647939}
\showDOI{\tempurl}


\bibitem[Guyon et~al\mbox{.}(2008)]%
        {guyon2008feature}
\bibfield{author}{\bibinfo{person}{Isabelle Guyon}, \bibinfo{person}{Steve
  Gunn}, \bibinfo{person}{Masoud Nikravesh}, {and} \bibinfo{person}{Lofti~A
  Zadeh}.} \bibinfo{year}{2008}\natexlab{}.
\newblock \bibinfo{booktitle}{\emph{Feature extraction: foundations and
  applications}}. Vol.~\bibinfo{volume}{207}.
\newblock \bibinfo{publisher}{Springer}.
\newblock
\urldef\tempurl%
\url{https://eprints.soton.ac.uk/261922/}
\showURL{%
\tempurl}


\bibitem[Hao et~al\mbox{.}(2013)]%
        {hao2013learning}
\bibfield{author}{\bibinfo{person}{Wangli Hao}, \bibinfo{person}{Jianwu Li},
  {and} \bibinfo{person}{Xiao Zhang}.} \bibinfo{year}{2013}\natexlab{}.
\newblock \showarticletitle{{Learning KPCA for face recognition}}. In
  \bibinfo{booktitle}{\emph{Communications in Computer and Information
  Science}}, Vol.~\bibinfo{volume}{375}. Springer, \bibinfo{pages}{142--146}.
\newblock
\showISBNx{9783642396779}
\showISSN{18650929}
\urldef\tempurl%
\url{https://doi.org/10.1007/978-3-642-39678-6_24}
\showDOI{\tempurl}


\bibitem[Helton and Davis(2003)]%
        {helton2003latin}
\bibfield{author}{\bibinfo{person}{J.C. Helton} {and} \bibinfo{person}{F.J.
  Davis}.} \bibinfo{year}{2003}\natexlab{}.
\newblock \showarticletitle{Latin hypercube sampling and the propagation of
  uncertainty in analyses of complex systems}.
\newblock \bibinfo{journal}{\emph{Reliability Engineering \& System Safety}}
  \bibinfo{volume}{81}, \bibinfo{number}{1} (\bibinfo{year}{2003}),
  \bibinfo{pages}{23--69}.
\newblock
\showISSN{0951-8320}
\urldef\tempurl%
\url{https://doi.org/10.1016/S0951-8320(03)00058-9}
\showDOI{\tempurl}


\bibitem[Herodotou et~al\mbox{.}(2020)]%
        {herodotou2020survey}
\bibfield{author}{\bibinfo{person}{Herodotos Herodotou},
  \bibinfo{person}{Yuxing Chen}, {and} \bibinfo{person}{Jiaheng Lu}.}
  \bibinfo{year}{2020}\natexlab{}.
\newblock \showarticletitle{A Survey on Automatic Parameter Tuning for Big Data
  Processing Systems}.
\newblock \bibinfo{journal}{\emph{ACM Comput. Surv.}} \bibinfo{volume}{53},
  \bibinfo{number}{2}, Article \bibinfo{articleno}{43} (\bibinfo{date}{apr}
  \bibinfo{year}{2020}), \bibinfo{numpages}{37}~pages.
\newblock
\showISSN{0360-0300}
\urldef\tempurl%
\url{https://doi.org/10.1145/3381027}
\showDOI{\tempurl}


\bibitem[Hindman et~al\mbox{.}(2011)]%
        {hindman2011mesos}
\bibfield{author}{\bibinfo{person}{Benjamin Hindman}, \bibinfo{person}{Andy
  Konwinski}, \bibinfo{person}{Matei Zaharia}, \bibinfo{person}{Ali Ghodsi},
  \bibinfo{person}{Anthony~D. Joseph}, \bibinfo{person}{Randy Katz},
  \bibinfo{person}{Scott Shenker}, {and} \bibinfo{person}{Ion Stoica}.}
  \bibinfo{year}{2011}\natexlab{}.
\newblock \showarticletitle{Mesos: A Platform for Fine-Grained Resource Sharing
  in the Data Center}. In \bibinfo{booktitle}{\emph{Proceedings of the 8th
  USENIX Conference on Networked Systems Design and Implementation}} (Boston,
  MA) \emph{(\bibinfo{series}{NSDI'11})}. \bibinfo{publisher}{USENIX
  Association}, \bibinfo{address}{USA}, \bibinfo{pages}{295–308}.
\newblock
\urldef\tempurl%
\url{https://dl.acm.org/doi/abs/10.5555/1972457.1972488}
\showURL{%
\tempurl}


\bibitem[Hoffman et~al\mbox{.}(2011)]%
        {hoffman2011portfolio}
\bibfield{author}{\bibinfo{person}{Matthew Hoffman}, \bibinfo{person}{Eric
  Brochu}, \bibinfo{person}{Nando de Freitas}, {et~al\mbox{.}}}
  \bibinfo{year}{2011}\natexlab{}.
\newblock \showarticletitle{Portfolio Allocation for Bayesian Optimization.}.
  In \bibinfo{booktitle}{\emph{UAI}}. Citeseer, \bibinfo{pages}{327--336}.
\newblock
\urldef\tempurl%
\url{https://dl.acm.org/doi/abs/10.5555/3020548.3020587}
\showURL{%
\tempurl}


\bibitem[Hsu et~al\mbox{.}(2018)]%
        {hsu2018arrow}
\bibfield{author}{\bibinfo{person}{Chin-Jung Hsu}, \bibinfo{person}{Vivek
  Nair}, \bibinfo{person}{Vincent~W. Freeh}, {and} \bibinfo{person}{Tim
  Menzies}.} \bibinfo{year}{2018}\natexlab{}.
\newblock \showarticletitle{Arrow: Low-Level Augmented Bayesian Optimization
  for Finding the Best Cloud VM}. In \bibinfo{booktitle}{\emph{2018 IEEE 38th
  International Conference on Distributed Computing Systems (ICDCS)}}.
  \bibinfo{pages}{660--670}.
\newblock
\urldef\tempurl%
\url{https://doi.org/10.1109/ICDCS.2018.00070}
\showDOI{\tempurl}


\bibitem[Huang et~al\mbox{.}(2010)]%
        {hibench}
\bibfield{author}{\bibinfo{person}{Shengsheng Huang}, \bibinfo{person}{Jie
  Huang}, \bibinfo{person}{Jinquan Dai}, \bibinfo{person}{Tao Xie}, {and}
  \bibinfo{person}{Bo Huang}.} \bibinfo{year}{2010}\natexlab{}.
\newblock \showarticletitle{The HiBench benchmark suite: Characterization of
  the MapReduce-based data analysis}. In \bibinfo{booktitle}{\emph{2010 IEEE
  26th International Conference on Data Engineering Workshops (ICDEW 2010)}}.
  \bibinfo{pages}{41--51}.
\newblock
\urldef\tempurl%
\url{https://doi.org/10.1109/ICDEW.2010.5452747}
\showDOI{\tempurl}


\bibitem[Ivanov and Beer(2015)]%
        {ivanov2015evaluating}
\bibfield{author}{\bibinfo{person}{Todor Ivanov} {and}
  \bibinfo{person}{Max-Georg Beer}.} \bibinfo{year}{2015}\natexlab{}.
\newblock \showarticletitle{{Evaluating Hive and Spark SQL with BigBench}}.
\newblock \bibinfo{journal}{\emph{arXiv preprint arXiv:1512.08417}}
  (\bibinfo{year}{2015}).
\newblock
\showeprint[arxiv]{1512.08417}
\urldef\tempurl%
\url{http://arxiv.org/abs/1512.08417}
\showURL{%
\tempurl}


\bibitem[Jia et~al\mbox{.}(2016)]%
        {jia2016auto}
\bibfield{author}{\bibinfo{person}{Zhen Jia}, \bibinfo{person}{Chao Xue},
  \bibinfo{person}{Guancheng Chen}, \bibinfo{person}{Jianfeng Zhan},
  \bibinfo{person}{Lixin Zhang}, \bibinfo{person}{Yonghua Lin}, {and}
  \bibinfo{person}{Peter Hofstee}.} \bibinfo{year}{2016}\natexlab{}.
\newblock \showarticletitle{Auto-tuning Spark big data workloads on POWER8:
  Prediction-based dynamic SMT threading}. In \bibinfo{booktitle}{\emph{2016
  International Conference on Parallel Architecture and Compilation Techniques
  (PACT)}}. \bibinfo{pages}{387--400}.
\newblock
\urldef\tempurl%
\url{https://doi.org/10.1145/2967938.2967957}
\showDOI{\tempurl}


\bibitem[Jones et~al\mbox{.}(1998)]%
        {jones1998efficient}
\bibfield{author}{\bibinfo{person}{Donald~R. Jones}, \bibinfo{person}{Matthias
  Schonlau}, {and} \bibinfo{person}{William~J. Welch}.}
  \bibinfo{year}{1998}\natexlab{}.
\newblock \showarticletitle{{Efficient Global Optimization of Expensive
  Black-Box Functions}}.
\newblock \bibinfo{journal}{\emph{Journal of Global Optimization}}
  \bibinfo{volume}{13}, \bibinfo{number}{4} (\bibinfo{year}{1998}),
  \bibinfo{pages}{455--492}.
\newblock
\showISSN{09255001}
\urldef\tempurl%
\url{https://doi.org/10.1023/A:1008306431147}
\showDOI{\tempurl}


\bibitem[Karimian-Aliabadi et~al\mbox{.}(2019)]%
        {karimian2019scalable}
\bibfield{author}{\bibinfo{person}{Soroush Karimian-Aliabadi},
  \bibinfo{person}{Danilo Ardagna}, \bibinfo{person}{Reza Entezari-Maleki},
  {and} \bibinfo{person}{Ali Movaghar}.} \bibinfo{year}{2019}\natexlab{}.
\newblock \showarticletitle{Scalable Performance Modeling and Evaluation of
  MapReduce Applications}. In \bibinfo{booktitle}{\emph{High-Performance
  Computing and Big Data Analysis}}, \bibfield{editor}{\bibinfo{person}{Lucio
  Grandinetti}, \bibinfo{person}{Seyedeh~Leili Mirtaheri}, {and}
  \bibinfo{person}{Reza Shahbazian}} (Eds.). \bibinfo{publisher}{Springer
  International Publishing}, \bibinfo{address}{Cham},
  \bibinfo{pages}{441--458}.
\newblock
\showISBNx{978-3-030-33495-6}
\urldef\tempurl%
\url{https://doi.org/10.1007/978-3-030-33495-6_34}
\showDOI{\tempurl}


\bibitem[Kunjir and Babu(2020)]%
        {kunjir2020black}
\bibfield{author}{\bibinfo{person}{Mayuresh Kunjir} {and}
  \bibinfo{person}{Shivnath Babu}.} \bibinfo{year}{2020}\natexlab{}.
\newblock \showarticletitle{Black or White? How to Develop an AutoTuner for
  Memory-Based Analytics}. In \bibinfo{booktitle}{\emph{Proceedings of the 2020
  ACM SIGMOD International Conference on Management of Data}} (Portland, OR,
  USA) \emph{(\bibinfo{series}{SIGMOD '20})}. \bibinfo{publisher}{Association
  for Computing Machinery}, \bibinfo{address}{New York, NY, USA},
  \bibinfo{pages}{1667–1683}.
\newblock
\showISBNx{9781450367356}
\urldef\tempurl%
\url{https://doi.org/10.1145/3318464.3380591}
\showDOI{\tempurl}


\bibitem[Li et~al\mbox{.}(2018)]%
        {li2019qtune}
\bibfield{author}{\bibinfo{person}{Guoliang Li}, \bibinfo{person}{Xuanhe Zhou},
  \bibinfo{person}{Shifu Li}, {and} \bibinfo{person}{Bo Gao}.}
  \bibinfo{year}{2018}\natexlab{}.
\newblock \showarticletitle{{QTune: A QueryAware database tuning system with
  deep reinforcement learning}}.
\newblock \bibinfo{journal}{\emph{Proceedings of the VLDB Endowment}}
  \bibinfo{volume}{12}, \bibinfo{number}{12} (\bibinfo{year}{2018}),
  \bibinfo{pages}{2118--2130}.
\newblock
\showISSN{21508097}
\urldef\tempurl%
\url{https://doi.org/10.14778/3352063.3352129}
\showDOI{\tempurl}


\bibitem[Lorenzo et~al\mbox{.}(2015)]%
        {lorenzo2015coefficient}
\bibfield{author}{\bibinfo{person}{Jose~Carlos Lorenzo},
  \bibinfo{person}{Lourdes Yabor}, \bibinfo{person}{Norma Medina},
  \bibinfo{person}{Nicolas Quintana}, {and} \bibinfo{person}{Vanessa Wells}.}
  \bibinfo{year}{2015}\natexlab{}.
\newblock \showarticletitle{Coefficient of variation can identify the most
  important effects of experimental treatments}.
\newblock \bibinfo{journal}{\emph{Notulae Botanicae Horti Agrobotanici
  Cluj-Napoca}} \bibinfo{volume}{43}, \bibinfo{number}{1}
  (\bibinfo{year}{2015}), \bibinfo{pages}{287--291}.
\newblock
\urldef\tempurl%
\url{https://www.notulaebotanicae.ro/index.php/nbha/article/view/9881/7790}
\showURL{%
\tempurl}


\bibitem[Lv et~al\mbox{.}(2015)]%
        {lv2015olap}
\bibfield{author}{\bibinfo{person}{Yanfei Lv}, \bibinfo{person}{Huihong He},
  \bibinfo{person}{Yasong Zheng}, \bibinfo{person}{Zhe Liu}, {and}
  \bibinfo{person}{Hong Zhang}.} \bibinfo{year}{2015}\natexlab{}.
\newblock \showarticletitle{OLAP query performance tuning in Spark}. In
  \bibinfo{booktitle}{\emph{Third International Conference on Cyberspace
  Technology (CCT 2015)}}. \bibinfo{pages}{1--5}.
\newblock
\urldef\tempurl%
\url{https://doi.org/10.1049/cp.2015.0832}
\showDOI{\tempurl}


\bibitem[Lv et~al\mbox{.}(2018)]%
        {counterminer}
\bibfield{author}{\bibinfo{person}{Yirong Lv}, \bibinfo{person}{Bin Sun},
  \bibinfo{person}{Qingyi Luo}, \bibinfo{person}{Jing Wang},
  \bibinfo{person}{Zhibin Yu}, {and} \bibinfo{person}{Xuehai Qian}.}
  \bibinfo{year}{2018}\natexlab{}.
\newblock \showarticletitle{Counterminer: Mining big performance data from
  hardware counters}. In \bibinfo{booktitle}{\emph{2018 51st Annual IEEE/ACM
  International Symposium on Microarchitecture (MICRO)}}. IEEE,
  \bibinfo{pages}{613--626}.
\newblock
\urldef\tempurl%
\url{https://doi.org/10.1109/MICRO.2018.00056}
\showDOI{\tempurl}


\bibitem[Marmolin(1986)]%
        {marmolin1986subjective}
\bibfield{author}{\bibinfo{person}{Hans Marmolin}.}
  \bibinfo{year}{1986}\natexlab{}.
\newblock \showarticletitle{{Subjective Mse Measures.}}
\newblock \bibinfo{journal}{\emph{IEEE Transactions on Systems, Man and
  Cybernetics}} \bibinfo{volume}{SMC-16}, \bibinfo{number}{3}
  (\bibinfo{year}{1986}), \bibinfo{pages}{486--489}.
\newblock
\showISSN{00189472}
\urldef\tempurl%
\url{https://doi.org/10.1109/tsmc.1986.4308985}
\showDOI{\tempurl}


\bibitem[Mika et~al\mbox{.}(1998)]%
        {mika1998kernel}
\bibfield{author}{\bibinfo{person}{Sebastian Mika}, \bibinfo{person}{Bernhard
  Sch\"{o}lkopf}, \bibinfo{person}{Alex Smola}, \bibinfo{person}{Klaus-Robert
  M\"{u}ller}, \bibinfo{person}{Matthias Scholz}, {and} \bibinfo{person}{Gunnar
  R\"{a}tsch}.} \bibinfo{year}{1998}\natexlab{}.
\newblock \showarticletitle{Kernel PCA and De-Noising in Feature Spaces}. In
  \bibinfo{booktitle}{\emph{Advances in Neural Information Processing
  Systems}}, \bibfield{editor}{\bibinfo{person}{M.~Kearns},
  \bibinfo{person}{S.~Solla}, {and} \bibinfo{person}{D.~Cohn}} (Eds.),
  Vol.~\bibinfo{volume}{11}. \bibinfo{publisher}{MIT Press}.
\newblock
\urldef\tempurl%
\url{https://proceedings.neurips.cc/paper/1998/file/226d1f15ecd35f784d2a20c3ecf56d7f-Paper.pdf}
\showURL{%
\tempurl}


\bibitem[Mockus(2012)]%
        {mockus2012bayesian}
\bibfield{author}{\bibinfo{person}{Jonas Mockus}.}
  \bibinfo{year}{2012}\natexlab{}.
\newblock \bibinfo{booktitle}{\emph{Bayesian approach to global optimization:
  theory and applications}}. Vol.~\bibinfo{volume}{37}.
\newblock \bibinfo{publisher}{Springer Science \& Business Media}.
\newblock
\urldef\tempurl%
\url{https://doi.org/10.1007/978-94-009-0909-0}
\showDOI{\tempurl}


\bibitem[Pavlo et~al\mbox{.}(2009)]%
        {pavlo2009comparison}
\bibfield{author}{\bibinfo{person}{Andrew Pavlo}, \bibinfo{person}{Erik
  Paulson}, \bibinfo{person}{Alexander Rasin}, \bibinfo{person}{Daniel~J.
  Abadi}, \bibinfo{person}{David~J. DeWitt}, \bibinfo{person}{Samuel Madden},
  {and} \bibinfo{person}{Michael Stonebraker}.}
  \bibinfo{year}{2009}\natexlab{}.
\newblock \showarticletitle{A Comparison of Approaches to Large-Scale Data
  Analysis}. In \bibinfo{booktitle}{\emph{Proceedings of the 2009 ACM SIGMOD
  International Conference on Management of Data}} (Providence, Rhode Island,
  USA) \emph{(\bibinfo{series}{SIGMOD '09})}. \bibinfo{publisher}{Association
  for Computing Machinery}, \bibinfo{address}{New York, NY, USA},
  \bibinfo{pages}{165–178}.
\newblock
\showISBNx{9781605585512}
\urldef\tempurl%
\url{https://doi.org/10.1145/1559845.1559865}
\showDOI{\tempurl}


\bibitem[Petridis et~al\mbox{.}(2017)]%
        {Petridis2017Trial}
\bibfield{author}{\bibinfo{person}{Panagiotis Petridis},
  \bibinfo{person}{Anastasios Gounaris}, {and} \bibinfo{person}{Jordi Torres}.}
  \bibinfo{year}{2017}\natexlab{}.
\newblock \showarticletitle{Spark Parameter Tuning via Trial-and-Error}. In
  \bibinfo{booktitle}{\emph{Advances in Big Data}},
  \bibfield{editor}{\bibinfo{person}{Plamen Angelov}, \bibinfo{person}{Yannis
  Manolopoulos}, \bibinfo{person}{Lazaros Iliadis}, \bibinfo{person}{Asim Roy},
  {and} \bibinfo{person}{Marley Vellasco}} (Eds.). \bibinfo{publisher}{Springer
  International Publishing}, \bibinfo{address}{Cham},
  \bibinfo{pages}{226--237}.
\newblock
\showISBNx{978-3-319-47898-2}
\urldef\tempurl%
\url{https://doi.org/10.1007/978-3-319-47898-2_24}
\showDOI{\tempurl}


\bibitem[Prats et~al\mbox{.}(2020)]%
        {prats2020you}
\bibfield{author}{\bibinfo{person}{David~Buchaca Prats},
  \bibinfo{person}{Felipe~Albuquerque Portella}, \bibinfo{person}{Carlos~H.A.
  Costa}, {and} \bibinfo{person}{Josep~Lluis Berral}.}
  \bibinfo{year}{2020}\natexlab{}.
\newblock \showarticletitle{{You only Run Once: Spark Auto-Tuning from a Single
  Run}}.
\newblock \bibinfo{journal}{\emph{IEEE Transactions on Network and Service
  Management}} \bibinfo{volume}{17}, \bibinfo{number}{4}
  (\bibinfo{year}{2020}), \bibinfo{pages}{2039--2051}.
\newblock
\showISSN{19324537}
\urldef\tempurl%
\url{https://doi.org/10.1109/TNSM.2020.3034824}
\showDOI{\tempurl}


\bibitem[Ramdane et~al\mbox{.}(2019)]%
        {ramdane2018partitioning}
\bibfield{author}{\bibinfo{person}{Yassine Ramdane}, \bibinfo{person}{Omar
  Boussaid}, \bibinfo{person}{Nadia Kabachi}, {and} \bibinfo{person}{Fadila
  Bentayeb}.} \bibinfo{year}{2019}\natexlab{}.
\newblock \showarticletitle{{Partitioning and Bucketing Techniques to Speed up
  Query Processing in Spark-SQL}}. In \bibinfo{booktitle}{\emph{Proceedings of
  the International Conference on Parallel and Distributed Systems - ICPADS}},
  Vol.~\bibinfo{volume}{2018-Decem}. IEEE, \bibinfo{pages}{142--151}.
\newblock
\showISBNx{9781538673089}
\showISSN{15219097}
\urldef\tempurl%
\url{https://doi.org/10.1109/PADSW.2018.8644891}
\showDOI{\tempurl}


\bibitem[Rasmussen(2004)]%
        {rasmussen2003gaussian}
\bibfield{author}{\bibinfo{person}{Carl~Edward Rasmussen}.}
  \bibinfo{year}{2004}\natexlab{}.
\newblock \showarticletitle{{Gaussian Processes in machine learning}}. In
  \bibinfo{booktitle}{\emph{Lecture Notes in Computer Science (including
  subseries Lecture Notes in Artificial Intelligence and Lecture Notes in
  Bioinformatics)}}, Vol.~\bibinfo{volume}{3176}. Springer,
  \bibinfo{pages}{63--71}.
\newblock
\showISSN{16113349}
\urldef\tempurl%
\url{https://doi.org/10.1007/978-3-540-28650-9_4}
\showDOI{\tempurl}


\bibitem[Rida(2020)]%
        {qayyum2020roadmap}
\bibfield{author}{\bibinfo{person}{Qayyum Rida}.}
  \bibinfo{year}{2020}\natexlab{}.
\newblock \showarticletitle{{A Roadmap Towards Big Data Opportunities, Emerging
  Issues and Hadoop as a Solution}}.
\newblock \bibinfo{journal}{\emph{International Journal of Education and
  Management Engineering}} \bibinfo{volume}{10}, \bibinfo{number}{4}
  (\bibinfo{year}{2020}), \bibinfo{pages}{8--17}.
\newblock
\showISSN{23053623}
\urldef\tempurl%
\url{https://doi.org/10.5815/ijeme.2020.04.02}
\showDOI{\tempurl}


\bibitem[Sethi et~al\mbox{.}(2019)]%
        {sethi2019presto}
\bibfield{author}{\bibinfo{person}{Raghav Sethi}, \bibinfo{person}{Martin
  Traverso}, \bibinfo{person}{Dain Sundstrom}, \bibinfo{person}{David
  Phillips}, \bibinfo{person}{Wenlei Xie}, \bibinfo{person}{Yutian Sun},
  \bibinfo{person}{Nezih Yegitbasi}, \bibinfo{person}{Haozhun Jin},
  \bibinfo{person}{Eric Hwang}, \bibinfo{person}{Nileema Shingte}, {and}
  \bibinfo{person}{Christopher Berner}.} \bibinfo{year}{2019}\natexlab{}.
\newblock \showarticletitle{{Presto: SQL on everything}}. In
  \bibinfo{booktitle}{\emph{Proceedings - International Conference on Data
  Engineering}}, Vol.~\bibinfo{volume}{2019-April}. IEEE,
  \bibinfo{pages}{1802--1813}.
\newblock
\showISBNx{9781538674741}
\showISSN{10844627}
\urldef\tempurl%
\url{https://doi.org/10.1109/ICDE.2019.00196}
\showDOI{\tempurl}


\bibitem[Shahriari et~al\mbox{.}(2015)]%
        {shahriari2015taking}
\bibfield{author}{\bibinfo{person}{Bobak Shahriari}, \bibinfo{person}{Kevin
  Swersky}, \bibinfo{person}{Ziyu Wang}, \bibinfo{person}{Ryan~P. Adams}, {and}
  \bibinfo{person}{Nando de Freitas}.} \bibinfo{year}{2015}\natexlab{}.
\newblock \showarticletitle{Taking the Human Out of the Loop: A Review of
  Bayesian Optimization}.
\newblock \bibinfo{journal}{\emph{Proc. IEEE}} \bibinfo{volume}{104},
  \bibinfo{number}{1} (\bibinfo{year}{2015}), \bibinfo{pages}{148--175}.
\newblock
\urldef\tempurl%
\url{https://doi.org/10.1109/JPROC.2015.2494218}
\showDOI{\tempurl}


\bibitem[Singhal and Singh(2018)]%
        {singhal2017performance}
\bibfield{author}{\bibinfo{person}{Rekha Singhal} {and}
  \bibinfo{person}{Praveen Singh}.} \bibinfo{year}{2018}\natexlab{}.
\newblock \showarticletitle{Performance Assurance Model for Applications on
  SPARK Platform}. In \bibinfo{booktitle}{\emph{Performance Evaluation and
  Benchmarking for the Analytics Era}},
  \bibfield{editor}{\bibinfo{person}{Raghunath Nambiar} {and}
  \bibinfo{person}{Meikel Poess}} (Eds.). \bibinfo{publisher}{Springer
  International Publishing}, \bibinfo{address}{Cham},
  \bibinfo{pages}{131--146}.
\newblock
\showISBNx{978-3-319-72401-0}
\urldef\tempurl%
\url{https://doi.org/10.1007/978-3-319-72401-0_10}
\showDOI{\tempurl}


\bibitem[Snoek et~al\mbox{.}(2012)]%
        {snoek2012practical}
\bibfield{author}{\bibinfo{person}{Jasper Snoek}, \bibinfo{person}{Hugo
  Larochelle}, {and} \bibinfo{person}{Ryan~P Adams}.}
  \bibinfo{year}{2012}\natexlab{}.
\newblock \showarticletitle{Practical Bayesian Optimization of Machine Learning
  Algorithms}. In \bibinfo{booktitle}{\emph{Advances in Neural Information
  Processing Systems}}, \bibfield{editor}{\bibinfo{person}{F.~Pereira},
  \bibinfo{person}{C.~J.~C. Burges}, \bibinfo{person}{L.~Bottou}, {and}
  \bibinfo{person}{K.~Q. Weinberger}} (Eds.), Vol.~\bibinfo{volume}{25}.
  \bibinfo{publisher}{Curran Associates, Inc.}
\newblock
\urldef\tempurl%
\url{https://proceedings.neurips.cc/paper/2012/file/05311655a15b75fab86956663e1819cd-Paper.pdf}
\showURL{%
\tempurl}


\bibitem[SparkConf(2022)]%
        {SparkConfWebsite}
SparkConf \bibinfo{year}{2022}\natexlab{}.
\newblock \bibinfo{booktitle}{\emph{Configuration - Spark 3.2.1
  Documentation}}.
\newblock
\urldef\tempurl%
\url{https://spark.apache.org/docs/latest/configuration.html}
\showURL{%
Retrieved March 28, 2022 from \tempurl}


\bibitem[SparkTuning(2022)]%
        {SparkConfTuningWebsite}
SparkTuning \bibinfo{year}{2022}\natexlab{}.
\newblock \bibinfo{booktitle}{\emph{Tuning - Spark 3.2.1 Documentation}}.
\newblock
\urldef\tempurl%
\url{https://spark.apache.org/docs/latest/tuning.html}
\showURL{%
Retrieved March 28, 2022 from \tempurl}


\bibitem[Suraweera and Mitrovic(2002)]%
        {suraweera2002kermit}
\bibfield{author}{\bibinfo{person}{Pramuditha Suraweera} {and}
  \bibinfo{person}{Antonija Mitrovic}.} \bibinfo{year}{2002}\natexlab{}.
\newblock \showarticletitle{KERMIT: A Constraint-Based Tutor for Database
  Modeling}. In \bibinfo{booktitle}{\emph{Intelligent Tutoring Systems}},
  \bibfield{editor}{\bibinfo{person}{Stefano~A. Cerri}, \bibinfo{person}{Guy
  Gouard{\`e}res}, {and} \bibinfo{person}{F{\`a}bio Paragua{\c{c}}u}} (Eds.).
  \bibinfo{publisher}{Springer Berlin Heidelberg}, \bibinfo{address}{Berlin,
  Heidelberg}, \bibinfo{pages}{377--387}.
\newblock
\showISBNx{978-3-540-47987-1}
\urldef\tempurl%
\url{https://doi.org/10.1007/3-540-47987-2_41}
\showDOI{\tempurl}


\bibitem[Sutton and Barto(1998)]%
        {sutton2018reinforcement}
\bibfield{author}{\bibinfo{person}{R.~S. Sutton} {and} \bibinfo{person}{A.~G.
  Barto}.} \bibinfo{year}{1998}\natexlab{}.
\newblock \showarticletitle{Reinforcement Learning: An Introduction}.
\newblock \bibinfo{journal}{\emph{Trans. Neur. Netw.}} \bibinfo{volume}{9},
  \bibinfo{number}{5} (\bibinfo{date}{sep} \bibinfo{year}{1998}),
  \bibinfo{pages}{1054}.
\newblock
\showISSN{1045-9227}
\urldef\tempurl%
\url{https://doi.org/10.1109/TNN.1998.712192}
\showDOI{\tempurl}


\bibitem[Thusoo et~al\mbox{.}(2009)]%
        {hive}
\bibfield{author}{\bibinfo{person}{Ashish Thusoo}, \bibinfo{person}{Joydeep~Sen
  Sarma}, \bibinfo{person}{Namit Jain}, \bibinfo{person}{Zheng Shao},
  \bibinfo{person}{Prasad Chakka}, \bibinfo{person}{Suresh Anthony},
  \bibinfo{person}{Hao Liu}, \bibinfo{person}{Pete Wyckoff}, {and}
  \bibinfo{person}{Raghotham Murthy}.} \bibinfo{year}{2009}\natexlab{}.
\newblock \showarticletitle{Hive: A Warehousing Solution over a Map-Reduce
  Framework}.
\newblock \bibinfo{journal}{\emph{Proc. VLDB Endow.}} \bibinfo{volume}{2},
  \bibinfo{number}{2} (\bibinfo{date}{aug} \bibinfo{year}{2009}),
  \bibinfo{pages}{1626–1629}.
\newblock
\showISSN{2150-8097}
\urldef\tempurl%
\url{https://doi.org/10.14778/1687553.1687609}
\showDOI{\tempurl}


\bibitem[{TPC Benchmark DS}(2020)]%
        {tpcds}
\bibfield{author}{\bibinfo{person}{{TPC Benchmark DS}}.}
  \bibinfo{year}{2020}\natexlab{}.
\newblock
  \bibinfo{howpublished}{\url{http://www.tpc.org/tpc\_documents\_current\_versions\\/pdf/tpc-ds_v2.3.0.pdf}}.
\newblock


\bibitem[Vavilapalli et~al\mbox{.}(2013)]%
        {vavilapalli2013yarn}
\bibfield{author}{\bibinfo{person}{Vinod~Kumar Vavilapalli},
  \bibinfo{person}{Arun~C. Murthy}, \bibinfo{person}{Chris Douglas},
  \bibinfo{person}{Sharad Agarwal}, \bibinfo{person}{Mahadev Konar},
  \bibinfo{person}{Robert Evans}, \bibinfo{person}{Thomas Graves},
  \bibinfo{person}{Jason Lowe}, \bibinfo{person}{Hitesh Shah},
  \bibinfo{person}{Siddharth Seth}, \bibinfo{person}{Bikas Saha},
  \bibinfo{person}{Carlo Curino}, \bibinfo{person}{Owen O'Malley},
  \bibinfo{person}{Sanjay Radia}, \bibinfo{person}{Benjamin Reed}, {and}
  \bibinfo{person}{Eric Baldeschwieler}.} \bibinfo{year}{2013}\natexlab{}.
\newblock \showarticletitle{Apache Hadoop YARN: Yet Another Resource
  Negotiator}. In \bibinfo{booktitle}{\emph{Proceedings of the 4th Annual
  Symposium on Cloud Computing}} (Santa Clara, California)
  \emph{(\bibinfo{series}{SOCC '13})}. \bibinfo{publisher}{Association for
  Computing Machinery}, \bibinfo{address}{New York, NY, USA}, Article
  \bibinfo{articleno}{5}, \bibinfo{numpages}{16}~pages.
\newblock
\showISBNx{9781450324281}
\urldef\tempurl%
\url{https://doi.org/10.1145/2523616.2523633}
\showDOI{\tempurl}


\bibitem[Vaz et~al\mbox{.}(2017)]%
        {vaz2017classification}
\bibfield{author}{\bibinfo{person}{Marcos Andr{\'{e}}~Braz Vaz},
  \bibinfo{person}{Paulo~Santana Pacheco}, \bibinfo{person}{Enio~J{\'{u}}nior
  Seidel}, {and} \bibinfo{person}{Angela~Pellegrin Ansuj}.}
  \bibinfo{year}{2017}\natexlab{}.
\newblock \showarticletitle{{Classification of the coefficient of variation to
  variables in beef cattle experiments}}.
\newblock \bibinfo{journal}{\emph{Ci{\^{e}}ncia Rural}} \bibinfo{volume}{47},
  \bibinfo{number}{11} (\bibinfo{year}{2017}).
\newblock
\urldef\tempurl%
\url{https://doi.org/10.1590/0103-8478cr20160946}
\showDOI{\tempurl}


\bibitem[Venkataraman et~al\mbox{.}(2016)]%
        {venkataraman2016ernest}
\bibfield{author}{\bibinfo{person}{Shivaram Venkataraman},
  \bibinfo{person}{Zongheng Yang}, \bibinfo{person}{Michael Franklin},
  \bibinfo{person}{Benjamin Recht}, {and} \bibinfo{person}{Ion Stoica}.}
  \bibinfo{year}{2016}\natexlab{}.
\newblock \showarticletitle{Ernest: Efficient Performance Prediction for
  {Large-Scale} Advanced Analytics}. In \bibinfo{booktitle}{\emph{13th USENIX
  Symposium on Networked Systems Design and Implementation (NSDI 16)}}.
  \bibinfo{publisher}{USENIX Association}, \bibinfo{address}{Santa Clara, CA},
  \bibinfo{pages}{363--378}.
\newblock
\showISBNx{978-1-931971-29-4}
\urldef\tempurl%
\url{https://www.usenix.org/conference/nsdi16/technical-sessions/presentation/venkataraman}
\showURL{%
\tempurl}


\bibitem[Wang et~al\mbox{.}(2016)]%
        {wang2016novel}
\bibfield{author}{\bibinfo{person}{Guolu Wang}, \bibinfo{person}{Jungang Xu},
  {and} \bibinfo{person}{Ben He}.} \bibinfo{year}{2016}\natexlab{}.
\newblock \showarticletitle{A Novel Method for Tuning Configuration Parameters
  of Spark Based on Machine Learning}. In \bibinfo{booktitle}{\emph{2016 IEEE
  18th International Conference on High Performance Computing and
  Communications; IEEE 14th International Conference on Smart City; IEEE 2nd
  International Conference on Data Science and Systems (HPCC/SmartCity/DSS)}}.
  \bibinfo{pages}{586--593}.
\newblock
\urldef\tempurl%
\url{https://doi.org/10.1109/HPCC-SmartCity-DSS.2016.0088}
\showDOI{\tempurl}


\bibitem[Wang and Khan(2015)]%
        {wang2015performance}
\bibfield{author}{\bibinfo{person}{Kewen Wang} {and} \bibinfo{person}{Mohammad
  Maifi~Hasan Khan}.} \bibinfo{year}{2015}\natexlab{}.
\newblock \showarticletitle{Performance Prediction for Apache Spark Platform}.
  In \bibinfo{booktitle}{\emph{2015 IEEE 17th International Conference on High
  Performance Computing and Communications, 2015 IEEE 7th International
  Symposium on Cyberspace Safety and Security, and 2015 IEEE 12th International
  Conference on Embedded Software and Systems}}. \bibinfo{pages}{166--173}.
\newblock
\urldef\tempurl%
\url{https://doi.org/10.1109/HPCC-CSS-ICESS.2015.246}
\showDOI{\tempurl}


\bibitem[Williams and Barber(1998)]%
        {williams1998bayesian}
\bibfield{author}{\bibinfo{person}{Christopher~K.I. Williams} {and}
  \bibinfo{person}{David Barber}.} \bibinfo{year}{1998}\natexlab{}.
\newblock \showarticletitle{{Bayesian classification with gaussian processes}}.
\newblock \bibinfo{journal}{\emph{IEEE Transactions on Pattern Analysis and
  Machine Intelligence}} \bibinfo{volume}{20}, \bibinfo{number}{12}
  (\bibinfo{year}{1998}), \bibinfo{pages}{1342--1351}.
\newblock
\showISSN{01628828}
\urldef\tempurl%
\url{https://doi.org/10.1109/34.735807}
\showDOI{\tempurl}


\bibitem[Yu et~al\mbox{.}(2018)]%
        {yu2018datasize}
\bibfield{author}{\bibinfo{person}{Zhibin Yu}, \bibinfo{person}{Zhendong Bei},
  {and} \bibinfo{person}{Xuehai Qian}.} \bibinfo{year}{2018}\natexlab{}.
\newblock \showarticletitle{Datasize-Aware High Dimensional Configurations
  Auto-Tuning of In-Memory Cluster Computing}.
\newblock \bibinfo{journal}{\emph{SIGPLAN Not.}} \bibinfo{volume}{53},
  \bibinfo{number}{2} (\bibinfo{date}{mar} \bibinfo{year}{2018}),
  \bibinfo{pages}{564–577}.
\newblock
\showISSN{0362-1340}
\urldef\tempurl%
\url{https://doi.org/10.1145/3296957.3173187}
\showDOI{\tempurl}


\bibitem[Zacheilas et~al\mbox{.}(2017)]%
        {zacheilas2017dione}
\bibfield{author}{\bibinfo{person}{Nikos Zacheilas}, \bibinfo{person}{Stathis
  Maroulis}, {and} \bibinfo{person}{Vana Kalogeraki}.}
  \bibinfo{year}{2017}\natexlab{}.
\newblock \showarticletitle{{Dione: Profiling spark applications exploiting
  graph similarity}}. In \bibinfo{booktitle}{\emph{Proceedings - 2017 IEEE
  International Conference on Big Data, Big Data 2017}},
  Vol.~\bibinfo{volume}{2018-Janua}. \bibinfo{pages}{389--394}.
\newblock
\showISBNx{9781538627143}
\urldef\tempurl%
\url{https://doi.org/10.1109/BigData.2017.8257950}
\showDOI{\tempurl}


\bibitem[Zaharia et~al\mbox{.}(2010)]%
        {spark-original}
\bibfield{author}{\bibinfo{person}{Matei Zaharia}, \bibinfo{person}{Mosharaf
  Chowdhury}, \bibinfo{person}{Michael~J. Franklin}, \bibinfo{person}{Scott
  Shenker}, {and} \bibinfo{person}{Ion Stoica}.}
  \bibinfo{year}{2010}\natexlab{}.
\newblock \showarticletitle{Spark: Cluster Computing with Working Sets}. In
  \bibinfo{booktitle}{\emph{Proceedings of the 2nd USENIX Conference on Hot
  Topics in Cloud Computing}} (Boston, MA)
  \emph{(\bibinfo{series}{HotCloud'10})}. \bibinfo{publisher}{USENIX
  Association}, \bibinfo{address}{USA}, \bibinfo{pages}{10}.
\newblock
\urldef\tempurl%
\url{https://dl.acm.org/doi/abs/10.5555/1863103.1863113}
\showURL{%
\tempurl}


\bibitem[Zar(2005)]%
        {zar2005spearman}
\bibfield{author}{\bibinfo{person}{Jerrold~H. Zar}.}
  \bibinfo{year}{2005}\natexlab{}.
\newblock \bibinfo{booktitle}{\emph{Spearman Rank Correlation}}.
\newblock \bibinfo{publisher}{John Wiley \& Sons, Ltd}.
\newblock
\showISBNx{9780470011812}
\urldef\tempurl%
\url{https://doi.org/10.1002/0470011815.b2a15150}
\showDOI{\tempurl}


\bibitem[Zebari et~al\mbox{.}(2020)]%
        {zebari2020comprehensive}
\bibfield{author}{\bibinfo{person}{Rizgar Zebari}, \bibinfo{person}{Adnan
  Abdulazeez}, \bibinfo{person}{Diyar Zeebaree}, \bibinfo{person}{Dilovan
  Zebari}, {and} \bibinfo{person}{Jwan Saeed}.}
  \bibinfo{year}{2020}\natexlab{}.
\newblock \showarticletitle{A comprehensive review of dimensionality reduction
  techniques for feature selection and feature extraction}.
\newblock \bibinfo{journal}{\emph{Journal of Applied Science and Technology
  Trends}} \bibinfo{volume}{1}, \bibinfo{number}{2} (\bibinfo{year}{2020}),
  \bibinfo{pages}{56--70}.
\newblock
\urldef\tempurl%
\url{https://doi.org/10.38094/jastt1224}
\showDOI{\tempurl}


\bibitem[Zhang et~al\mbox{.}(2019)]%
        {zhang2019end}
\bibfield{author}{\bibinfo{person}{Ji Zhang}, \bibinfo{person}{Yu Liu},
  \bibinfo{person}{Ke Zhou}, \bibinfo{person}{Guoliang Li},
  \bibinfo{person}{Zhili Xiao}, \bibinfo{person}{Bin Cheng},
  \bibinfo{person}{Jiashu Xing}, \bibinfo{person}{Yangtao Wang},
  \bibinfo{person}{Tianheng Cheng}, \bibinfo{person}{Li Liu},
  \bibinfo{person}{Minwei Ran}, {and} \bibinfo{person}{Zekang Li}.}
  \bibinfo{year}{2019}\natexlab{}.
\newblock \showarticletitle{{An end-to-end automatic cloud database tuning
  system using deep reinforcement learning}}. In
  \bibinfo{booktitle}{\emph{Proceedings of the ACM SIGMOD International
  Conference on Management of Data}}. \bibinfo{pages}{415--432}.
\newblock
\showISBNx{9781450356435}
\showISSN{07308078}
\urldef\tempurl%
\url{https://doi.org/10.1145/3299869.3300085}
\showDOI{\tempurl}


\bibitem[Zhang and Yang(2016)]%
        {zhang2016big}
\bibfield{author}{\bibinfo{person}{Tonglin Zhang} {and}
  \bibinfo{person}{Baijian Yang}.} \bibinfo{year}{2016}\natexlab{}.
\newblock \showarticletitle{{Big Data Dimension Reduction Using PCA}}. In
  \bibinfo{booktitle}{\emph{Proceedings - 2016 IEEE International Conference on
  Smart Cloud, SmartCloud 2016}}. IEEE, \bibinfo{pages}{152--157}.
\newblock
\showISBNx{9781509052622}
\urldef\tempurl%
\url{https://doi.org/10.1109/SmartCloud.2016.33}
\showDOI{\tempurl}


\bibitem[Zhu et~al\mbox{.}(2017)]%
        {zhu2017bestconfig}
\bibfield{author}{\bibinfo{person}{Yuqing Zhu}, \bibinfo{person}{Jianxun Liu},
  \bibinfo{person}{Mengying Guo}, \bibinfo{person}{Yungang Bao},
  \bibinfo{person}{Wenlong Ma}, \bibinfo{person}{Zhuoyue Liu},
  \bibinfo{person}{Kunpeng Song}, {and} \bibinfo{person}{Yingchun Yang}.}
  \bibinfo{year}{2017}\natexlab{}.
\newblock \showarticletitle{{BestConfig: Tapping the performance potential of
  systems via automatic configuration tuning}}. In
  \bibinfo{booktitle}{\emph{SoCC 2017 - Proceedings of the 2017 Symposium on
  Cloud Computing}} \emph{(\bibinfo{series}{SoCC '17})}.
  \bibinfo{publisher}{Association for Computing Machinery},
  \bibinfo{address}{New York, NY, USA}, \bibinfo{pages}{338--350}.
\newblock
\showISBNx{9781450350280}
\urldef\tempurl%
\url{https://doi.org/10.1145/3127479.3128605}
\showDOI{\tempurl}
\showeprint[arxiv]{1710.03439}


\bibitem[Álvaro Brandón~Hernández et~al\mbox{.}(2018)]%
        {hernandez2018using}
\bibfield{author}{\bibinfo{person}{Álvaro Brandón~Hernández},
  \bibinfo{person}{María~S. Perez}, \bibinfo{person}{Smrati Gupta}, {and}
  \bibinfo{person}{Victor Muntés-Mulero}.} \bibinfo{year}{2018}\natexlab{}.
\newblock \showarticletitle{Using machine learning to optimize parallelism in
  big data applications}.
\newblock \bibinfo{journal}{\emph{Future Generation Computer Systems}}
  \bibinfo{volume}{86} (\bibinfo{year}{2018}), \bibinfo{pages}{1076--1092}.
\newblock
\showISSN{0167-739X}
\urldef\tempurl%
\url{https://doi.org/10.1016/j.future.2017.07.003}
\showDOI{\tempurl}


\end{thebibliography}

\end{document}